\documentclass[fleqn,usenatbib]{mnras}

\usepackage{newtxtext,newtxmath}

\usepackage[T1]{fontenc}

\DeclareRobustCommand{\VAN}[3]{#2}
\let\VANthebibliography\thebibliography
\def\thebibliography{\DeclareRobustCommand{\VAN}[3]{##3}\VANthebibliography}

\usepackage{graphicx}	
\usepackage{amsmath}	
\usepackage{dcolumn}
\usepackage{bm}
\usepackage{hyperref}
\usepackage{color}
\usepackage{xcolor, soul}
\sethlcolor{orange} 

\newcommand{\bna}{\bm \nabla}
\newcommand{\bal}{\bm \alpha}
\newcommand{\be}{\begin{equation}}
\newcommand{\ee}{\end{equation}}
\newcommand{\ba}{\begin{eqnarray}}
\newcommand{\ea}{\end{eqnarray}}

\newcommand{\en}{\nonumber\\}
\newcommand{\de}{\delta}

\newcommand{\xx}{\mathbf{x}}
\newcommand{\rr}{\mathbf{r}}
\newcommand{\uu}{\mathbf{u}}

\newcommand{\ab}{{\bf A}}
\newcommand{\ktil}{{\tilde{\kappa}}}
\newcommand{\dif}{{\rm d}}

\title[Anisotropic strong lensing in SIDM and WDM]{Anisotropic strong lensing as a probe of dark matter self-interactions}

\author[B. Dhanasingham et al.]{Birendra Dhanasingham$^{1}$\thanks{E-mail: birendradh@unm.edu},
Francis-Yan Cyr-Racine$^{1}$\thanks{E-mail: fycr@unm.edu}, 
Charlie Mace$^{2,3,4}$, Annika H. G. Peter$^{2,3,4,5}$, ~and~
\newauthor Andrew Benson$^{6}$
\\
$^{1}$Department of Physics and Astronomy, University of New Mexico, 210 Yale Blvd NE, Albuquerque, NM 87106, USA\\
$^{2}$Department of Physics, The Ohio State University, 191 W. Woodruff Ave., Columbus, OH 43210, USA\\
$^{3}$Center for Cosmology and Astroparticle Physics, The Ohio State University, 191 W. Woodruff Ave., Columbus, OH 43210, USA\\
$^{4}$Department of Astronomy, The Ohio State University, 140 W. 18th Ave., Columbus, OH 43210, USA\\
$^{5}$School of Natural Sciences, Institute for Advanced Study, 1 Einstein Drive, Princeton, NJ 08540, USA\\
$^{6}$Carnegie Observatories, 813 Santa Barbara Street, Pasadena, CA 91101, USA
}

\date{Accepted 2023 October 6. Received 2023 September 27; in original form 2023 June 26}

\pubyear{2023}

\begin{document}
\label{firstpage}
\pagerange{\pageref{firstpage}--\pageref{lastpage}}
\maketitle

\begin{abstract}

Galaxy-scale strongly lensed systems have been shown to provide a unique technique for exploring the underlying physics of dark matter at sub-galactic scales. In the past, much attention was given to detecting and studying individual haloes in a strong lens system. In addition to the subhaloes, line-of-sight haloes contribute significantly to the small perturbations in lensed images. In prior work, we demonstrated that these line-of-sight haloes imprint a distinctive anisotropic signature and hence give rise to a detectable non-zero parity-even quadrupole moment in the effective convergence field's two-point correlation function. In this study, we show that these line-of-sight haloes also produce a non-zero curl component of the effective deflection field with a parity-odd quadrupole moment of the two-point function. These multipole moments have the ability to statistically separate line-of-sight haloes from dark matter substructure. In this paper, we examine how these multipole moments evolve in the presence of warm dark matter and self-interacting dark matter in terms of central density evolution and dark matter halo abundance. Importantly, we show that these different multipole moments display exquisite sensitivity to both the amplitude and the velocity dependence of the dark matter self-interaction cross-section. Our approach opens the door for strong lensing observations to probe dark matter self-interaction over a broad range of relative velocities.


\end{abstract}



\section{Introduction}

Alternative models to the Lambda cold dark matter ($\Lambda$CDM) model have gained popularity as potential tools for understanding the $\Lambda$CDM model's inadequacies in the non-linear regime \citep[see e.g.,][]{Bullock:2017xww}. For instance, self-interacting dark matter (SIDM) \citep{Spergel_2000, Tulin:2012wi, Tulin:2013teo, Rocha_2013, Peter_2013, Kaplinghat_2013, Kaplinghat_2014_1, Kaplinghat_2014_2, Kaplinghat:2015aga, Elbert_2015, Burger_2019}, in which dark matter particles can transfer momentum and energy through scattering, is characterized by haloes with a different density profile as compared to CDM. This profile could be either cored if the interaction cross-section is not too large and haloes are not too evolved, or very dense and steep if the haloes enter the gravothermal-collapse regime \citep{Lynden-Bell_wood_1968, Balberg:2002ue, Balberg_2002, Ahn:2004xt, Kahlhoefer_2019, Nishikawa_2020, gilman2021strong, Zeng_2022, Meshveliani:2022rih, Yang:2022hkm, Yang:2023jwn, Yang_2023, Zhong:2023yzk}.

Another possibility is that warm dark matter (WDM), which deviates from the hierarchical structure formation of the $\Lambda$CDM picture, exhibits a tiny abundance of low-mass dark matter haloes because free streaming effects erase the small density perturbations below a characteristic length scale \citep{Bond_1983, Bode:2000gq, Dalcanton:2000hn, Schneider:2013ria, Viel_2013, Benson_2013, Pullen_2014, Lovell_2020}. The suppression of the abundance of WDM haloes and the core or cuspy central densities of SIDM haloes considerably impact the subtle perturbations in a strong lensed image, making galaxy-scale strong lens systems a powerful tool for studying different dark matter models.

Strong gravitational lensing is a promising method for probing the distribution of dark matter on sub-galactic scales and at redshifts higher than the Local Group of galaxies \citep[see e.g.,][]{Mao:1998aa, Chiba:aa, Dalal_2002, Metcalf:ac, Kochanek_2003, Keeton:2003aa, Kochanek_2004, koopmans2005}. A close analysis of these strongly lensed images \citep[see e.g.,][]{Bolton_2006, Bolton_2008, Gavazzi_2008, Auger_2009, Brownstein_2012, Shu_2016, oldham_2017, Cornachione_2018, DES:2019mte} reveals small localized gravitational perturbations beyond the smooth mass distribution of the main lens galaxy, which are mostly caused by the main lens substructure or line-of-sight haloes distributed between the observer and the source. A careful examination of these subtle perturbations has the potential to disclose the physics underneath these small-scale structures \citep{vegetti2009a, vegetti2009b, Xu:2011ru, vegetti2010a, vegetti2010b, vegetti2012, vegetti_2014, Li:2016afu, Hezaveh2016, minor2016, Gilman:2019vca, Gilman:2019nap, Gilman:2019bdm, Amorisco_2021, Minor_2021_2, Minor_2021, Sengul:2021lxe, Sengul:2022edu, Laroche_2022, Hogg_2023, Keeley:2023sad, Vegetti_2023_6}.

When considering the collective effect of multiple lens planes in a strong lens system, \cite{Dhanasingham_2022} pointed out that the dark matter substructure and line-of-sight haloes contribute differently to the two lensing potentials discussed under the effective multiplane gravitational lensing formalism. In this approach, the effective deflection field is decomposed into a gradient of an effective scalar potential and the curl of an effective vector potential, where the latter is zero under single-plane lensing. Many previous studies on the weak lensing by large-scale structure on cosmic microwave background (CMB) temperature and polarization fluctuations took a similar approach, pointing out a rotation of these fluctuation fields by going beyond the Born approximation and into the post-Born regime \citep{Cooray_2002, Hirata_2003, Cooray_2005, Robertson_2023}. In addition, \cite{Pen_2006} discussed the induced rotation of quadruply lensed sources in the presence of uncorrelated large-scale structure. In a strong lens system, the projected mass density derived by considering the curl of the effective deflection field is thus purely originating from line-of-sight objects, causing rotational distortion in the strongly lensed images.

According to \cite{Dhanasingham_2022}, unlike dark matter subhaloes, line-of-sight haloes leave distinctive anisotropic traces and quadrupole patterns in the two-dimensional projected mass density field. In this work, we measure the distinct quadrupole signature on the two-point function imprinted by these haloes in the presence of WDM and SIDM. Many earlier papers have proposed the use of angular-averaged two-point function monopole as a feasible tool for studying dark matter physics on sub-galactic scales \citep[see e.g.,][]{Hezaveh_2014, Chatterjee:2018ast, DiazRivero:2017xkd, DiazRivero:2018oxk, Brennan:2018jhq, Cyr-Racine_2019, CaganSengul:2020nat, Bayer_2018, Bayer_2023}. In addition to this monopole generated by both substructure and line-of-sight haloes, we focus for the first time on anisotropic imprints generated by isolated line-of-sight haloes with two different dark matter theories. This effort will be useful in testing and improving our understanding of dark matter microphysics in conjunction with many ground- and space-based surveys this decade to observe galaxy--galaxy strong lenses \citep{Serjeant_2014, Collett_2015, Serjeant_2017, Metcalf_2019, Weiner_2020, Mao_2022} and the development of machine learning algorithms to identify these strong lens systems \citep[see e.g.,][]{Hezaveh_2017, Perreault_Levasseur_2017, Schaefer_2018, Vernardos_2020, Cheng_2020, Li_2020, Ostdiek:2020cqz,Ostdiek:2020mvo, Legin_2022, Wilde_2022, Rezaei_2022, Zhang:2022djp, Adam_2023, Wagner-Carena_2023, Rojas_2023, Canameras_2023}. In particular, as we will show in this paper, these various multipole moments are sensitive to both the amplitude and velocity dependence of the dark matter self-interaction cross-section. Since the purpose of this work is to demonstrate how different dark matter scenarios change the amplitude and shape of two-point correlation function multipoles rather than analyse specific known lensed systems, we focus below on relatively simple lens configurations. However, we emphasize that important aspects of observed strong lens systems (e.g., macrolens model, baryon disruption effects, etc.) will need to be carefully modelled in any realistic analyses, as they can impact the shape and amplitude of the two-point correlation function. We leave such a detailed analysis to future work.

In this work, \textsc {pyhalo}\footnote{https://github.com/dangilman/pyHalo} \citep{gilman2021strong} is used to render dark matter substructure and line-of-sight haloes on top of the open-source \textsc{python} software package \textsc {lenstronomy}\footnote{https://github.com/lenstronomy} \citep{Birrer:2018xgm, Birrer2021} as the lensing computation tool. Based on the Planck 2018 \citep{planck_2018} results, we assume a flat $\rm \Lambda CDM$ cosmology throughout this work. Additionally, we set the redshift of the source to $z_{\rm s}=1.0$ and that of the primary lens to $z_{\rm macro}=0.5$. We take into account a power-law ellipsoid main lens mass profile with the Einstein radius $\theta_{\rm E}=1.0$ arcsec, the eccentricity components $(e_1, e_2)=(0.05, 0.08)$, and the logarithmic slope $\gamma = 2.078$ \citep{Auger_2010}. The components of the external gravitational shear field caused by matter in the neighbourhood of this main lens are set to $(\gamma_1, \gamma_2)=(0.01, -0.01)$. We utilize the mass definition of $M_{200}$ with regard to the critical density of the Universe at the halo redshift throughout this work.

This manuscript is organized as follows: In Section~\ref{Sec: Realizations}, we discuss how to generate dark matter halo realizations using the CDM, WDM, and SIDM frameworks, while accounting for free-streaming effects and dark matter self-interactions. Section~\ref{Sec: Eff Theory} introduces the fundamental ideas of effective multiplane lensing and various angular and symmetry structures in effective deflection fields, as well as methods for quantifying anisotropies and symmetries. In Section~\ref{Results}, we discuss the impact of WDM and SIDM on strong lensing anisotropy measurements. Finally, in Section~\ref{Conclusions}, we summarize our main findings and analyse the broad implications.
\section{Generating Dark Matter Halo Realizations} \label{Sec: Realizations}
This section describes how to generate dark matter substructure and line-of-sight halo realizations considering the CDM, SIDM, and WDM frameworks.

\subsection{Cold and warm dark matter haloes}\label{WDM} 

Here we discuss our dark matter substructure and line-of-sight density profile choices, taking into account both the CDM and WDM scenarios, as well as the suppression of WDM halo central densities due to the free-streaming effect of WDM particles.

\subsubsection{Halo density profiles}

The CDM subhaloes are modelled as truncated Navarro--Frenk--White (TNFW) profiles \citep{tNFW_2009} in our study, and the substructure is tidally truncated using a Roche limit approximation. Moreover, we use NFW \citep{NFW_1996} density profiles to simulate line-of-sight dark matter haloes and we compute the scale radii and densities using the mass--concentration relationship presented in \cite{Diemer:2018vmz} with a scatter in the concentration of 0.13 dex.

Unlike CDM haloes,  WDM haloes are subject to free-streaming effects, as discussed in the following section. This effect alters the central densities of TNFW subhaloes and the NFW line-of-sight halo mass profiles of WDM haloes, and therefore we employ a rescaled version of the CDM mass--concentration relationship for WDM haloes to account for this free-streaming effects, as discussed further below.

\subsubsection{Free-streaming and mass--concentration relation}

The characteristic non-negligible thermal velocities of WDM particles in the early Universe result in the free streaming of particles that erase small density perturbations in the matter distribution below the free-streaming length and inhibiting the formation of the structure below this length-scale, thus suppressing the linear matter power spectrum on small scales. The free-streaming length directly depends on the thermal velocity distribution of the dark matter particles and hence the WDM particle mass and formation mechanism \citep{Bond_1983, Schneider:2013ria, Benson_2013}.

The half-mode mass, $m_{\rm hm}$, which is defined in terms of the power spectrum wave number where the WDM transfer function relative to CDM is damped by 0.5, is frequently used to model the free-streaming effects on the WDM matter power spectrum. The abundance of WDM haloes is strongly suppressed below the half-mode mass as a result of free streaming, and WDM haloes also have a different internal structure than CDM haloes due to this effect. We discuss the free-streaming effects on the halo mass function in Section~\ref{Sec: Realizations}. In this work, we assume thermal relic WDM with particle mass $m_{\rm WDM}$ and thus use the relation between the half-mode mass and the dark matter mass $m_{\rm hm} \propto m_{\rm WDM}^{-3.33}$ \citep{Schneider_2012}. When this relationship is normalized using the scaling $2 \times 10^8{\rm M}_\odot h^{-1} \sim 3.3{\rm keV}$ constrained from the Lyman-$\alpha$ flux power spectrum measurements presented in \cite{Viel_2013}, the half-mode mass may be written down as
\be 
m_{\rm hm}(m_{\rm WDM})=10^{10}\left( \frac{m_{\rm WDM}}{1{\rm keV}} \right)^{-3.33}{\rm M}_\odot h^{-1}.
\ee

\begin{figure}
\begin{center}
\includegraphics[clip, trim=0cm 0cm 0cm 0cm, width=0.481\textwidth]{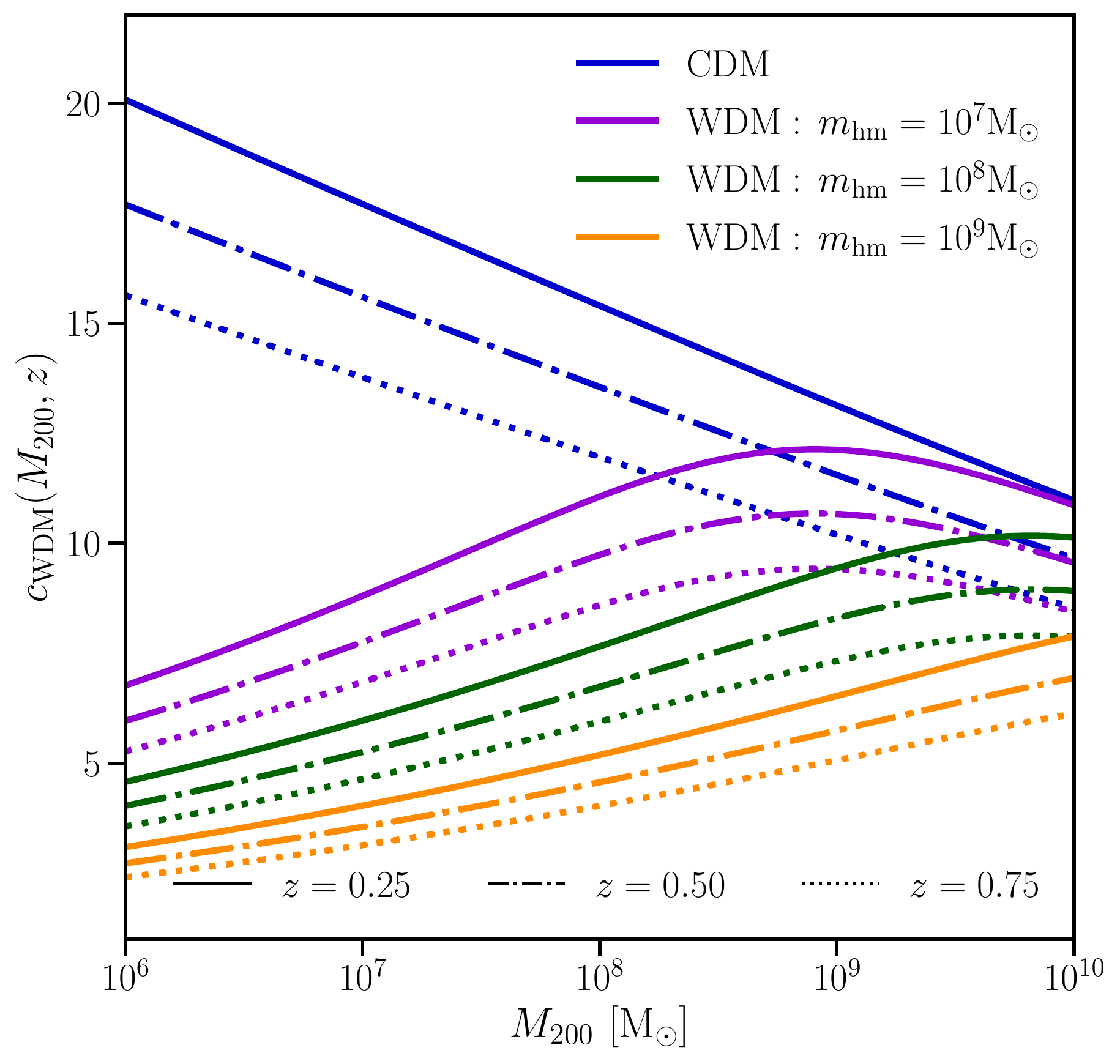}
\end{center}
\caption{\label{WDM_mass_con} The mass--concentration relation for CDM and WDM models as a function of half-mode mass ($m_{\rm hm}$) and the halo redshift ($z$).}
\end{figure}

In contrast to CDM haloes, the free-streaming effect limits the fine-grained phase space density of WDM haloes \citep{Maccio_2012, Shao_2013}, hence restricting the central density of these objects. Moreover, haloes below the half-mode mass form later in WDM models, when the Universe is less dense, and so have lower concentrations than in the CDM model. Because the subtle perturbations of lensed images caused by dark matter substructure and line-of-sight haloes are highly sensitive to dark matter halo central densities, suppression of halo concentration has an important effect on strong lensing observables. In this study, we implement the parametrization proposed by \cite{Bose:2015mga} to model the free-streaming effects on the mass--concentration relationship of WDM haloes, which has the form
\be \label{WDM_Concentration}
\frac{c_{\rm WDM}(m,z)}{c_{\rm CDM}(m,z)}=(1+z)^{\beta(z)}\left( 1+60 \frac{m_{\rm hm}}{m}\right)^{-0.17},
\ee where $\beta(z)=0.026z-0.04$. Here $c_{\rm CDM}(m,z)$ is the CDM mass--concentration model from \cite{Diemer:2018vmz} with a scatter of 0.13 dex. Fig.~\ref{WDM_mass_con} illustrates the mass--concentration relation for CDM and WDM models as a function of half-mode mass and the halo redshift. We observe that haloes with masses up to two orders of magnitude above $m_{\rm hm}$ have suppressed concentration in the WDM case. We incorporate this halo concentration suppression into our WDM models in order to derive more accurate dark matter predictions for strong lensing observables.

\subsection{Self-interacting dark matter haloes}

To generate the SIDM halo realizations, we utilize the model presented by \cite{gilman2021strong}. In this section, we present a brief overview of this SIDM model.

\subsubsection{Interaction cross-section}
Let us first consider the simple velocity-dependent cross-section introduced by \cite{gilman2021strong}, which is parameterized as
\be
\sigma(v) = \frac{\sigma_{0}}{\left(1+\frac{v^2}{v^2_0} \right)^2},
\ee where $v$ is the relative velocity of the dark matter particles, $\sigma_0$ is the normalization that determines the amplitude of the cross-section, and the $\sigma (v)$ behaves as $\sigma_v \propto v^{-4}$ beyond the characteristic velocity $v_0$. This choice of interaction cross-section  enables sizable scattering of dark matter particles at low relative velocities while avoiding the constraints on the cross-section at high velocities from galaxies and galaxy clusters \citep{Peter_2013, Kim_2017, Robertson_2019, Banerjee_2020, Sagunski_2021, McDaniel_2021, Andrade_2022, Eckert_2022, Cross_2023}.

\subsubsection{Core formation, core-collapse, and core-collapse probability}

The presence of self-interactions can dramatically affect the internal dynamics of haloes by allowing different dark matter particles to exchange energy and momentum. In a standard NFW profile, the temperature and line-of-sight velocity dispersion of dark matter particles is greater towards the halo's outskirts. The presence of self-interaction allows heat to flow in and dark matter particles to flow out until an isothermal density core is formed and the central velocity dispersion is equalized \citep{Ahn:2004xt, Vogelsberger_2012, Rocha_2013, Elbert_2015, Burger_2019}. This core has a high temperature relative to the halo outskirts, creating a gradient that now favours an outward heat flow. The core then begins to shrink while its density increases, due to the negative heat capacity of such a self-gravitating system. This rapid core collapse, also known as the gravothermal catastrophe, generates a sharp density cusp \citep{Lynden-Bell_wood_1968, Balberg:2002ue, Balberg_2002}.

According to \cite{Nishikawa_2020}, dark matter haloes subjected to tidal stripping exhibit accelerated core-collapse compared to isolated dark matter haloes. This faster core-collapse occurs after only $\mathcal{O}(10)t_0$, as opposed to the longer time-scales of $\mathcal{O}(100)t_0$ taken by isolated haloes, which did not experience any tidal forces during their evolution \citep{Kahlhoefer_2019, Nishikawa_2020, Sameie_2020, Correa_2021, Zeng_2022, Yang:2022hkm, Yang_2023}. The time-scale for structure evolution $t_0$ here is determined by the values of $\sigma_0$, scale density, and $v_0$. The substructure and line-of-sight halo time scales are therefore set to $t_{\rm sub}=10t_0$ and $t_{\rm LOS}=100t_0$, respectively. Given the time since the halo's collapse $t_{\rm halo}$, we adopt the probability of core-collapse introduced by \cite{gilman2021strong}, parameterized as:

\be 
P_{\rm collapse} =
  \begin{cases}
0  & \quad t_{\rm halo} \leq 0.5t',\\
\frac{t_{\rm halo}-t'}{t'}  & \quad t_{\rm halo} \leq 2t',\\
1  & \quad t_{\rm halo} > 2t',
  \end{cases}
\ee where $t'$ is either $t_{\rm LOS}$ or $t_{\rm sub}$. It should be noted that this model is still approximate and has not yet been calibrated by simulations.

\subsubsection{Density profiles of cored and core-collapsed haloes}

We model the main galaxy substructure and line-of-sight dark matter haloes before the core-collapse as cored truncated NFW (cored-TNFW) profiles as
\begin{equation} \label{ctnfw}
    \rho(x, \beta, \tau)=\frac{\rho_{\rm s}}{(x^a+\beta^a)^\frac{1}{a}(1+x)^2}\frac{\tau^2}{\tau^2+x^2}, 
\end{equation} where $x=r/r_{\rm s}$, $\tau=r_{\rm t}/r_{\rm s}$,  $\beta=r_{\rm c}/r_{\rm s}$, $r_{\rm t}$ is the truncation radius, $r_{\rm c}$ is the core radius, and $r_{\rm s}$ is the scale radius. Appendix~\ref{App. A} summarizes the procedure for computing the parameter $\beta$ discussed in \cite{gilman2021strong}. We set the value of $a$ to 10 and calculate the scale radius $r_{\rm s}$ and scale density $\rho_{\rm s}$ for an NFW profile using the mass--concentration relationship proposed by \cite{Diemer:2018vmz} with a scatter of 0.13 dex.

In order to model the steep central density cusps in core-collapsed dark matter haloes, we use the power-law density profile given by  
\begin{equation}
    \rho(r, r_{\rm c}, x_{\rm match})=\rho_{\rm c}(x_{\rm match})\left(1+\frac{r^2}{r_{\rm c}^2} \right)^{-\frac{\gamma}{2}},
\end{equation} where $\gamma$ is the logarithmic profile slope, $\rho_{\rm c}$ is the core density, and $r_{\rm c}$ is the core radius. The parameter $x_{\rm match}$ determines the density profile normalization $\rho_{\rm c}$ so that the enclosed mass of the NFW profile inside $x_{\rm match}r_{\rm s}$ matches the mass of the core-collapsed halo inside $x_{\rm match}r_{\rm s}$. In our model, the core radius $r_{\rm c}$ that keeps the total mass within $r_{\rm s}$ finite, is set to $0.05r_{\rm s}$ and $x_{\rm match}$ and $\gamma$ are set to 2.16 and 3.0, respectively.

\subsection{Subhalo and line-of-sight halo mass functions}
\subsubsection{Populating CDM and SIDM haloes}

Considering both the SIDM and the benchmark CDM models, we generate multiple random realizations of dark matter substructure and line-of-sight dark matter haloes with the masses of the haloes range from $10^6{\rm M_\odot}$ to $10^{10}{\rm M_\odot}$, and $M_{\rm halo} = 10^{13.3}{\rm M}_\odot$ being the main halo mass. Here we omit haloes with masses greater than $10^{10}{\rm M_\odot}$ from our realizations since these massive haloes are likely to have a baryonic mass component and so be observable or lead to a large enough lensing effect to be directly included in the main lens model \citep[see e.g.,][]{2019MNRAS.484.4726B}.

The subhaloes are populated using the power-law mass function given in equation~\eqref{Eq. 12} with a logarithmic slope of $\alpha = -1.90$ and a pivot mass of $m_0=10^{8}{\rm M_\odot}$. This value is compatible with previously predicted power-law slope values by $N$-body simulations \citep{Benson_2020, Springel_2008}.
Given the halo mass $m$ at infall and normalization $\Sigma_{\rm sub}=0.025 {\rm kpc^{-2}}$ \citep{Gilman:2019vca, Gilman:2019nap, Nadler:2021dft}, the subhalo mass function takes the form 

\be \label{Eq. 12}
\frac{{\rm d}^2N_{\rm sub, CDM}}{{\rm d}m{\rm d}A}= \frac{{\rm d}^2N_{\rm sub, SIDM}}{{\rm d}m{\rm d}A}=\frac{\Sigma_{\rm sub}}{m_0}\left(\frac{m}{m_0} \right)^\alpha\mathcal{F}(M_{\rm halo}, z).
\ee Here 
\be \label{Eq. 13}
\log_{10}(\mathcal{F}) = k_1 \log_{10}\left(\frac{M_{\rm halo}}{10^{13}\rm M_\odot}\right)+k_2 \log_{10}(z+0.5)
\ee
is the scaling function that determines the differential projected number density of the substructure as a function of host halo mass $M_{\rm halo}$ and redshift $z$, with $(k_1, k_2) = (0.88, 1.70)$ \citep{Gilman:2019bdm} predicted using the \textsc {Galacticus} semi-analytical model \citep{BENSON_galacticus, Pullen_2014, Yang_2020}. 

We render line-of-sight dark matter haloes in the previously specified mass range using the Sheth--Tormen (ST) mass function \citep{Sheth_2001} with two modifications proposed by \cite{Gilman:2019vca}. The first re-scaling factor $\delta_{\rm LOS}$  proposed by \cite{Gilman:2019vca} accounts for systematic fluctuations in the mean number of haloes predicted by the Sheth--Tormen mass function as well as baryonic effects on small-scale structure formation \citep{Gilman:2019bdm, Benson_2020}. The second factor $\xi_{\rm 2halo}$ is the two-halo term, which represents the influence of correlated structures in the vicinity of the host dark matter halo on the three-dimensional two-point correlation function \citep{Gilman:2019vca, gilman2021strong, Lazar_2021}. Then the CDM and SIDM line-of-sight halo mass functions have the form 
\begin{align} \label{Eq. 14}
   \frac{{\rm d}^2N_{\rm LOS, CDM}}{{\rm d}m{\rm d}V} & = \frac{{\rm d}^2N_{\rm LOS, SIDM}}{{\rm d}m{\rm d}V} \en
   &= \de_{\rm LOS}\left(1+\xi_{\rm 2halo}(r, M_{\rm halo}, z) \right)\left[\frac{{\rm d}^2N_{\rm}}{{\rm d}m{\rm d}V}\right]_{\rm ST},  
\end{align} where
\be 
\xi_{\rm 2halo}(r, M_{\rm halo}, z) = b(M_{\rm halo}, z)\, \xi_{\rm lin}(r, z).
\ee The term $\xi_{\rm 2halo}(r, M_{\rm halo}, z)$ is determined by the halo bias $b(M_{\rm halo}, z)$ around the parent halo with mass $M_{\rm halo}$ as computed in \cite{Sheth_torman_1999} and the linear matter correlation function $\xi_{\rm lin}(r, z)$ at a distance $r$ computed using the linear power spectrum at redshift $z$. Unless otherwise specified in the paper, the line-of-sight mass function normalization $\de_{\rm LOS}$ is set to 1.0. 

\subsubsection{WDM halo mass functions}

As discussed in Section~\ref{WDM}, free-streaming of fully thermalized WDM particles results in a strong suppression of the WDM halo abundance below the half-mode mass. As a result, over the course of this study, we employ WDM substructure and line-of-sight mass functions proposed by \cite{Lovell_2020}, which provide a reasonable fit to the simulation results. By adopting the suppression factors, which are dependent on the halo mass and the half-mode mass, the subhalo and line-of-sight halo mass functions are then given by

\be \label{WDM_mf}
\frac{\dif N_{\rm sub, WDM}}{\dif \log m}=\frac{\dif N_{\rm sub, CDM}}{\dif \log m}\left[1+\left(4.2\frac{m_{\rm hm}}{m} \right)^{2.5} \right]^{-0.2}
\ee and

\be \label{WDM_mf_los}
\frac{\dif N_{\rm LOS,WDM}}{\dif \log m}=\frac{\dif N_{\rm LOS,CDM}}{\dif \log m}\left[1+\left(2.3\frac{m_{\rm hm}}{m} \right)^{0.8} \right]^{-1.0},
\ee respectively. 

\begin{figure}
\begin{center}
\includegraphics[clip, trim=0cm 0cm 0cm 0cm, width=0.481\textwidth]{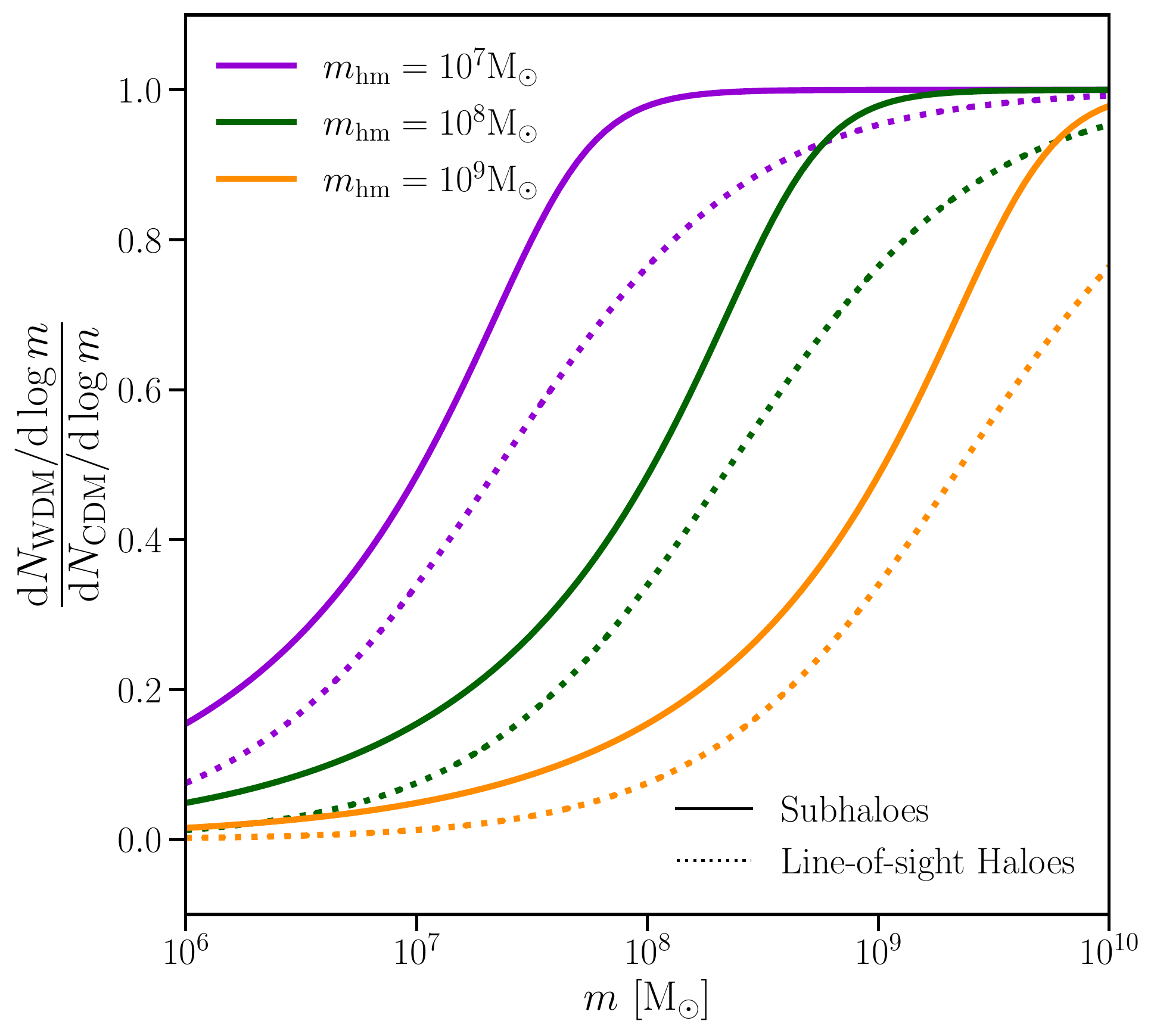}
\end{center}
\caption{\label{WDM_mass_func} The ratio of WDM to CDM mass functions as a function of half-mode mass for main lens substructure (using the bound mass definition) and line-of-sight dark matter haloes (using the mass definition $M_{200}$).}
\end{figure}

The ratio of the WDM and CDM mass functions obtained by equations~\eqref{WDM_mf} and \eqref{WDM_mf_los} as a function of half-mode mass ($m_{\rm hm}$) for subhaloes and line-of-sight haloes is shown in Fig.~\ref{WDM_mass_func}. In comparison to the CDM model, the WDM model exhibits a substantial suppression of the number of haloes below $m_{\rm hm}$ due to free streaming effects. The WDM subhalo mass function turnover occurs at lower masses than the WDM line-of-sight halo mass function, which is consistent with the fact that the bound subhalo mass definition we used for substructure accounts for tidal truncation, whereas the virial mass definition, $M_{200}$, we use for line-of-sight haloes is unaffected by such effects.

\section{Effective Multiplane Lensing and Different Angular and Symmetry Structures in Convergence Fields} \label{Sec: Eff Theory}
In this section, we look at the concept of effective multiplane gravitational lensing and the different anisotropic and symmetry structures produced by line-of-sight dark matter haloes in the divergence and the curl of the effective deflection field briefly.

\subsection{Effective multiplane lensing}

Photons encounter repeated deflections by dark matter haloes along the line-of-sight as they travel through the Universe from a background source to an observer. The usual method for dealing with this multiplane gravitational lensing involves a recursive lens equation that must be solved for each lens plane \citep{Blandford_1986, McCully_2014, Schneider_2014}. This usual strategy, however, is computationally expensive.

The effective multiplane lensing approach takes into account the cumulative influence of all the lens planes in the strong lens system and reduces it to a single mapping between the observer and the source \citep{Gilman:2019vca, Gilman:2019nap, Dhanasingham_2022}. The two-dimensional non-linear mapping from the source plane to the image plane in single-plane lensing is a pure gradient of a scalar lensing potential. However, due to the non-linear coupling between connective lens planes, this map no longer acts as a pure gradient of a scalar potential in the multiplane lensing phenomenon. To capture these non-linear coupling effects, the effective multiplane technique outlined in \cite{Dhanasingham_2022} incorporates a curl of an effective vector potential, $\ab_{\rm eff}$, in addition to the gradient of the effective scalar potential, $\phi_{\rm eff}$, and so gives the lens equation as 

\be\label{eq:main_lens_eq}
\uu (\xx) = \xx - \bal_{\rm eff}(\xx),  
\ee where the "effective" deflection field $\bal_{\rm eff}(\xx)$ is

\be \label{eq:alpha_eff_eq}
\bal_{\rm eff}(\xx) = \bna \phi_{\rm eff}(\xx) + \bna \times \ab_{\rm eff}(\xx).
\ee

\noindent Here, $\uu, \xx \in {\mathbb{R}^2}$ are the source and image plane coordinates, respectively. It is worth noting that the effective vector potential, $\ab_{\rm eff}$, points purely in the line-of-sight (or $\hat{\bf z}$) direction. Without any approximation, potential functions $\phi_{\rm eff}$ and $\ab_{\rm eff}$ encode the entire non-linear multiplane lensing map from the source plane to the image plane.

\begin{figure*}
\begin{center}
	\includegraphics[clip, trim=0cm 0cm 0cm 0cm, width=0.9\textwidth]{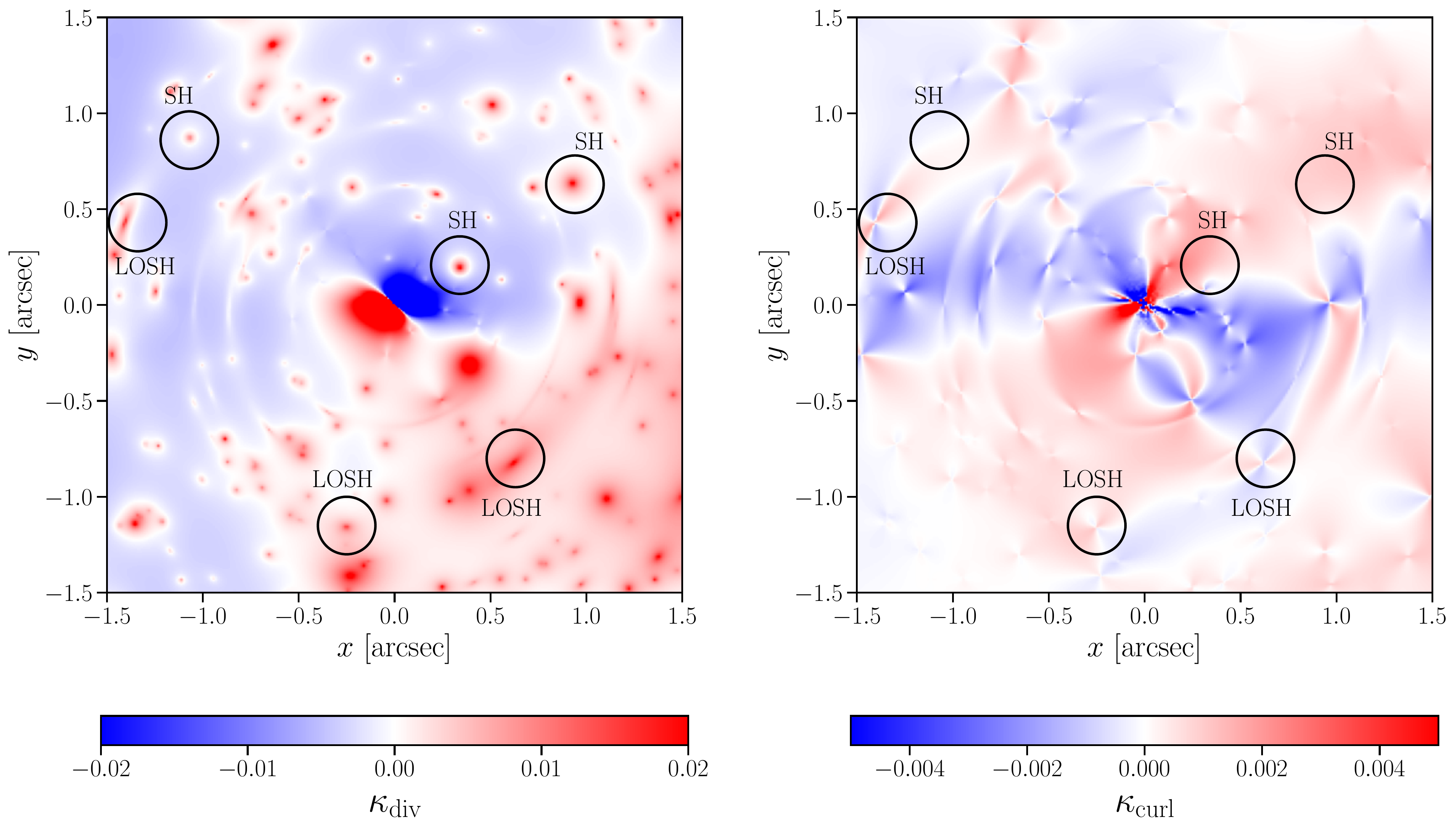}
\end{center}	
\caption{\label{fig:SH_LOSH_Fig} The $\kappa_{\rm div}$ (left panel) and $\kappa_{\rm curl}$ (right panel) convergence maps for a CDM halo realization include both the contribution from line-of-sight dark matter haloes (LOSH) and lens galaxy substructure (SH), with $(\Sigma_{\rm sub}/{\rm kpc^{-2}}, \de_{\rm LOS}, \log_{10}({m_{\rm min}}/{\rm M_\odot}), \log_{10}({m_{\rm max}}/{\rm M_\odot}))=(0.025, 1.0, 7, 10)$, taking into account the effective multiplane gravitational lensing. Take note of the varying angular structure in each case. Line-of-sight dark matter haloes seem warped in the tangential direction in the left panel, while subhaloes are usually circular. Also, line-of-sight haloes are accompanied by a quadrupolar pattern in the right panel, whereas subhaloes do not contribute to this map.}
\end{figure*}

As long as the vector field $\bal_{\rm eff}$ is twice continuously differentiable, the general decomposition introduced in equation~\eqref{eq:alpha_eff_eq} follows from the Helmholtz theorem. Potentials $\phi_{\rm eff}$ and $\ab_{\rm eff}$ can be computed for any behaviour of the effective deflection field at large distances from the centre of the lens by including the appropriate boundary terms.

It is useful to introduce the divergence ($\kappa_{\rm div}$) and curl ($\kappa_{\rm curl}$) of the effective deflection field in the study of multiple gravitational lensing, so that \citep{Gilman:2019vca, Dhanasingham_2022} 

\be \label{Eq. kappa div}
\kappa_{\rm div} \equiv \frac{1}{2} \bna \cdot \bal_{\rm eff} - \kappa_0 , 
\ee
and
\be \label{Eq. kappa curl}
\kappa_{\rm curl} \equiv \frac{1}{2} \bna \times \bal_{\rm eff} \cdot \hat{\bf z}.
\ee We subtract the projected mass density of the single-plane main lens model, $\kappa_0$, from the divergence of the effective deflection field to remove the dominant lensing contribution from the so-called macrolens. The $\kappa_{\rm curl}$ disappears in single-plane gravitational lensing because the effective deflection field is a pure gradient. Fig.~\ref{fig:SH_LOSH_Fig} depicts these two quantities for a CDM model.

\subsubsection{Convergence fields with different angular structures}

The different angular structures present in the $\kappa_{\rm div}$ and $\kappa_{\rm curl}$ convergence fields of a CDM halo realization containing both the line-of-sight haloes and the main lens substructure are depicted in Fig.~\ref{fig:SH_LOSH_Fig}. According to \cite{Dhanasingham_2022}, line-of-sight haloes appear stretched in the angular direction and form arc-like shapes in the $\kappa_{\rm div}$ map (left panel) because their deflection field is distorted by the main lens due to the non-linear nature of multiplane lensing. In addition to the obvious break in homogeneity caused by the macrolens, these arc-like structures add anisotropic signatures to the $\kappa_{\rm div}$ map, whereas subhaloes remain largely circular and thus isotropic. Of course, due to tidal effects, main-lens subhaloes could be elliptical or stretched in specific directions, but given the projected nature of lensing, we expect the ellipticity direction from those to be random rather than preferably aligned in the angular direction. By comparing the left and right panels of Fig.~\ref{fig:SH_LOSH_Fig}, it is also possible to see that the stretched features in the $\kappa_{\rm div}$ map correspond to quadrupolar patterns in the $\kappa_{\rm curl}$ map. The circular features in the $\kappa_{\rm div}$ field, which correspond to subhaloes, on the other hand, are not accompanied by any specific features in the $\kappa_{\rm curl}$ map.

 These apparent anisotropic signatures in the $\kappa_{\rm div}$ field and accompanying quadrupolar pattern in the $\kappa_{\rm curl}$ map, as stated by \cite{Dhanasingham_2022}, propose a way to distinguish between the subhaloes and line-of-sight dark matter haloes behind and in front of the main lens. In this paper, we exploit the diverse angular structures created by dark matter haloes to investigate several dark matter models.

\subsubsection{Parity and E-- and B-- modes in effective multiplane lensing} \label{Parity_E_and_B}

\begin{figure}
\begin{center}
\includegraphics[clip, trim=0cm 0cm 0cm 0cm, width=0.32\textwidth]{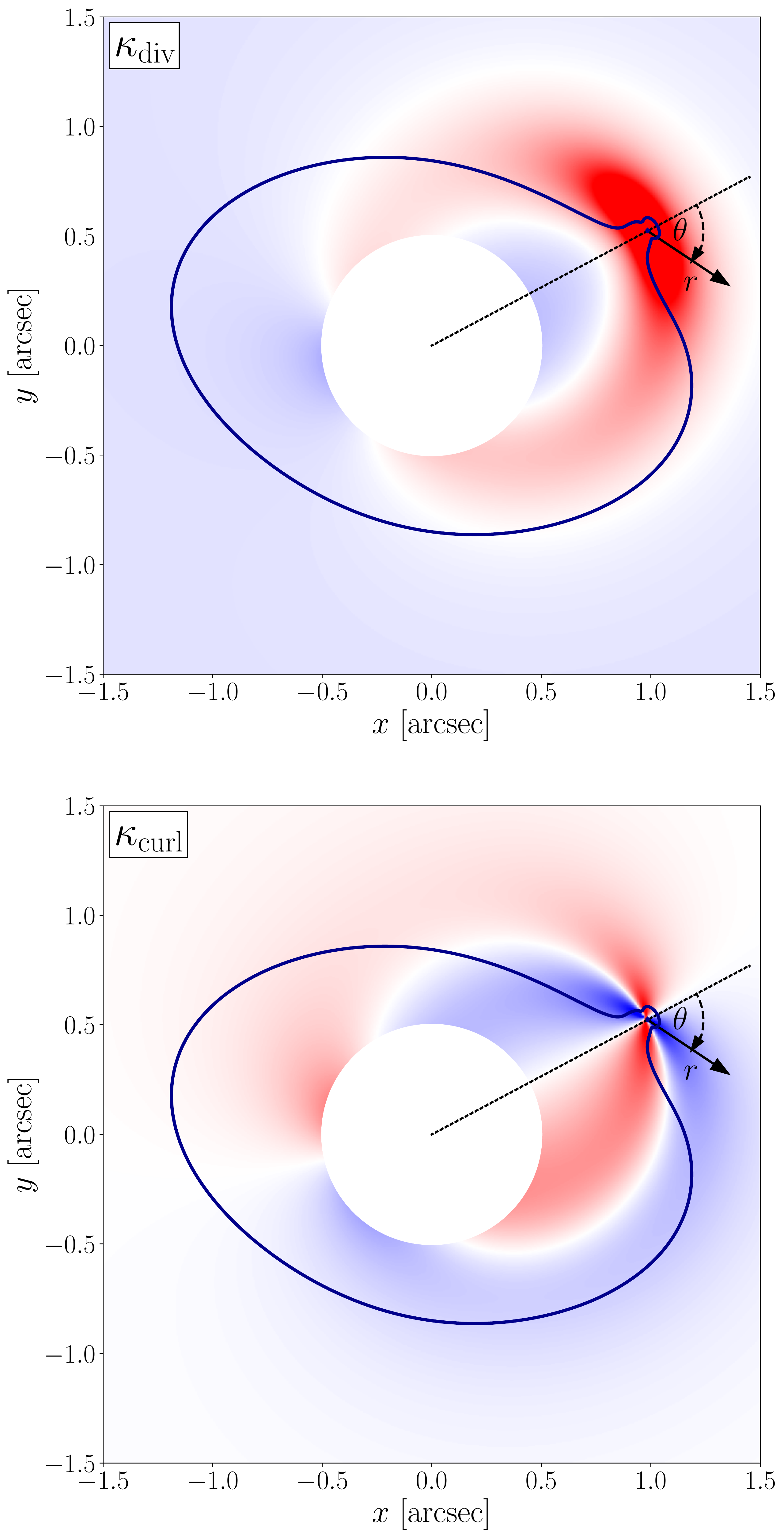}
\end{center}
\caption{\label{Parity}The distinct angular and symmetry structures are shown in the $\kappa_{\rm div}$ and $\kappa_{\rm curl}$ maps for a simple lens system with a NFW line-of-sight halo. The critical curve of the strong lens system is represented by the solid blue line.}
\end{figure}

Consider the arc-shaped structure in the $\kappa_{\rm div}$ convergence field and the accompanying quadrupole in the $\kappa_{\rm curl}$ field formed by a mass distribution along the line-of-sight. $\theta \in [0, \pi]$ is the angle measured in the clockwise direction from the centre of the projected mass density with respect to the line-of-sight, as shown in Fig.~\ref{Parity}. Indeed, in the $\kappa_{\rm div}$ and $\kappa_{\rm curl}$ maps, these line-of-sight haloes exhibit quadrupolar patterns with two different symmetry structures.

The transformation $\theta \rightarrow  -\theta$ gives $\kappa_{\rm div}(r,-\theta ) \approx \kappa_{\rm div}(r,\theta)$ and leaves the $\kappa_{\rm div}$ unchanged (see Fig.~\ref{Parity}). After a parity transformation, the $\kappa_{\rm div}$ is therefore locally invariant and exhibits \textit{even} parity. As a result, in effective multiplane lensing, these tangential distortions with even parity can be introduced as ``\textit{E-modes}''. Unlike the symmetry structure of the $\kappa_{\rm div}$ map, the $\kappa_{\rm curl}$ map transforms as $\kappa_{\rm curl}(r, -\theta) \approx -\kappa_{\rm curl}(r, \theta)$ and thus exhibits \textit{odd} parity under the transformation $\theta \rightarrow  -\theta$. Therefore, we referred to these unique quadrupoles with odd parity as ``\textit{B-modes}''. In this paper, we combine previously unstudied parity information found in the divergence and curl components of the convergence field of a strong lens system with their unique angular structures, including quadrupoles and anisotropies, to investigate various dark matter models using the theory of effective multiplane lensing.

\subsection{Quantifying anisotropies, quadrupoles, and symmetries}

In this subsection, we present methods for quantifying anisotropies and quadrupoles formed by line-of-sight haloes in $\kappa_{\rm div}$ and $\kappa_{\rm curl}$ convergence fields in order to extract useful information on the dark matter microphysics of line-of-sight haloes under the effect of mass-sheet degeneracy.

\subsubsection{Mass-sheet degeneracy}

In general, strong lensing systems have an inherent mathematical degeneracy in which a simultaneous transformation of the source and lens planes known as the source position transform \citep{Schneider:2014hsi, Unruh:2017yhi, Wertz_2018} can render certain astrometric and photometric lensing observables invariant. Because we observe strong lenses in angular projection on the sky and these angles are ratios of distances, we may always rescale such distances while keeping the observed angles constant. Several lensing-related works have addressed this mathematical degeneracy \citep[see e.g.,][]{Schneider_1995, Bradac_2004, Schneider_2013, Rexroth_2016, Birrer_2021, Cremonese_2021, Dhanasingham_2022}, and this can be broken only by observations on an absolute scale, such as the inherent size of the lensed source or the relative time delay between lensed images. 

The well-known mass-sheet degeneracy \citep{Falco_1985} is the scalar portion of this transformation. The mass-sheet transform (MST) converts the effective deflection field as:
\be \label{MST: alp_eff}
\bal '_{\rm eff}(\xx)=\lambda \bal_{\rm eff}(\xx) + (1-\lambda)\xx
\ee  by isotropically mapping the source plane coordinates as $\uu \rightarrow  \lambda\uu$. For this mapping, the effective convergence field, $\kappa_{\rm eff}=\frac{1}{2} \bna \cdot \bal_{\rm eff}$, can be rescaled as:
\be 
\kappa'_{\rm eff}(\xx) = \lambda\kappa_{\rm eff}(\xx)+(1-\lambda).
\ee We may then write down the transformation of the $\kappa_{\rm div}$ filed under MST as:
\be
\kappa'_{\rm div}=\lambda \kappa_{\rm div}
\ee using the MST of the macrolens convergence field, $\kappa_0'(\xx)=\lambda\kappa_0(\xx)+(1-\lambda)$. Similarly, we may rescale the $\kappa_{\rm curl}$ field as: 
\be
\kappa'_{\rm curl}=\lambda \kappa_{\rm curl}
\ee by considering the curl of the effective deflection field provided in equation~\eqref{MST: alp_eff}. Therefore, MST scales the two-point correlation function of a given convergence field as $\xi(\rr) \rightarrow \lambda^2\xi(\rr)$, and thereby biases its amplitude by a factor of $\lambda^2$. \cite{Dhanasingham_2022} used a normalized two-point correlation function approach to eliminate the effects of mass-sheet degeneracy, but for simplicity, we do not consider this normalization here because we are primarily interested in the overall shape of the two-point correlation function. Nonetheless, this normalization factor could be trivially restored to remove the impact of the MST.


\subsubsection{The two-point correlation function} \label{2PCF}

As stated in \cite{Dhanasingham_2022}, the contribution of the line-of-sight halo to the convergence field $\kappa_{\rm div}$ is no longer statistically isotropic, and there are also interesting quadrupole structures at the positions of the line-of-sight haloes present in the convergence field $\kappa_{\rm curl}$. To capture the fruitful information encoded in these anisotropies and quadrupole structures, we employ the image plane-averaged two-point correlation function, denoted as:

\be \label{eq:corr_func_def}
\xi(\rr)=\frac{1}{A}\int_A {\rm d}^2\rr_1 \,\, \big[\kappa(\rr_1) - \left<\kappa(\rr_1) \right> \big] \,\, \big[\kappa(\rr_2) - \left<\kappa(\rr_2)\right>\big],
\ee where $\rr_1$ and $\rr_2=\rr_1+\rr$ are the position vectors of two points from the center of the convergence field, $\rr$ is the vector linking these two points, and $A$ is the area of the image where the spatial average is performed. Here, $\kappa$ represents either $\kappa_{\rm div}$ or $\kappa_{\rm curl}$ and $\left<\kappa(\rr) \right>$ is the mean of the convergence map.

On the convergence map, we use a binary annular mask with the region of interest set to 1 and the domain to be masked out set to 0 to focus our attention on the relevant region close to the Einstein radius of the main lens. Using the point-wise multiplications of masks, $W(\rr)$, and convergence maps such that $\tilde{\kappa}(\rr)= W(\rr) \cdot \kappa(\rr)$, the masked correlation function of the convergence overdensity field \citep[see e.g.,][for similar works]{Padfield2010, Padfield2012,Dhanasingham_2022} can be written as:


\begin{align}\label{Eq.Num}
   \xi(\rr) & = \frac{1}{A} \int {\rm d}^2 \rr_1 \,\,  \ktil(\rr_1) \, \ktil(\rr+\rr_1) \en
    & \hspace{1cm} - \frac{\int {\rm d}^2 \rr_1 \,\,  \ktil(\rr_1) \, W(\rr+\rr_1) \cdot \int {\rm d}^2 \rr_1 \,\,  W(\rr_1) \, \ktil(\rr+\rr_1)}{A\int {\rm d}^2 \rr_1 \,\,  W(\rr_1) \, W(\rr+\rr_1)}.
\end{align}

\subsubsection{Decomposing the two-point function of $\kappa_{\rm div}$ field}

The distortions of the projected mass density of line-of-sight haloes in the tangential direction exhibit quadrupoles with even parity, as detailed in Section~\ref{Parity_E_and_B}. We decompose the two-point function of $\kappa_{\rm div}$ onto an orthonormal basis of Chebyshev polynomials, $T_\ell(\cos \theta)=\cos (\ell \theta)$, as:
\be \label{Eq.corr_dec}
\xi_{\rm div}(\rr)=\xi_{\rm div}(r, \theta)=\sum_{\ell=0}^\infty \xi_{\rm div, \ell}(r)\cos (\ell \theta),
\ee  to characterize the significant information hidden in these anisotropies together with their even parity. Here $\theta$ is the angle between $\rr$ and the line-of-sight vector from the centre to the mid-point of the vector $\rr$. In equation~\eqref{Eq.corr_dec}, the term $\xi_{\rm div,\ell}(r)$ denotes the correlation multipole of order $\ell$, which is thus given by
\be 
\xi_{\rm div, \ell}(r) = \frac{2-\de_{\ell 0}}{\pi}\int_{0}^\pi {\rm d} \theta \,\, \xi_{\rm div}(r, \theta)\cos (\ell \theta),
\ee
 where $\de_{ij}$ is the Kronecker delta. 

\begin{figure*}
\begin{center}
	\includegraphics[clip, trim=0cm 0cm 0cm 0cm, width=0.8\textwidth]{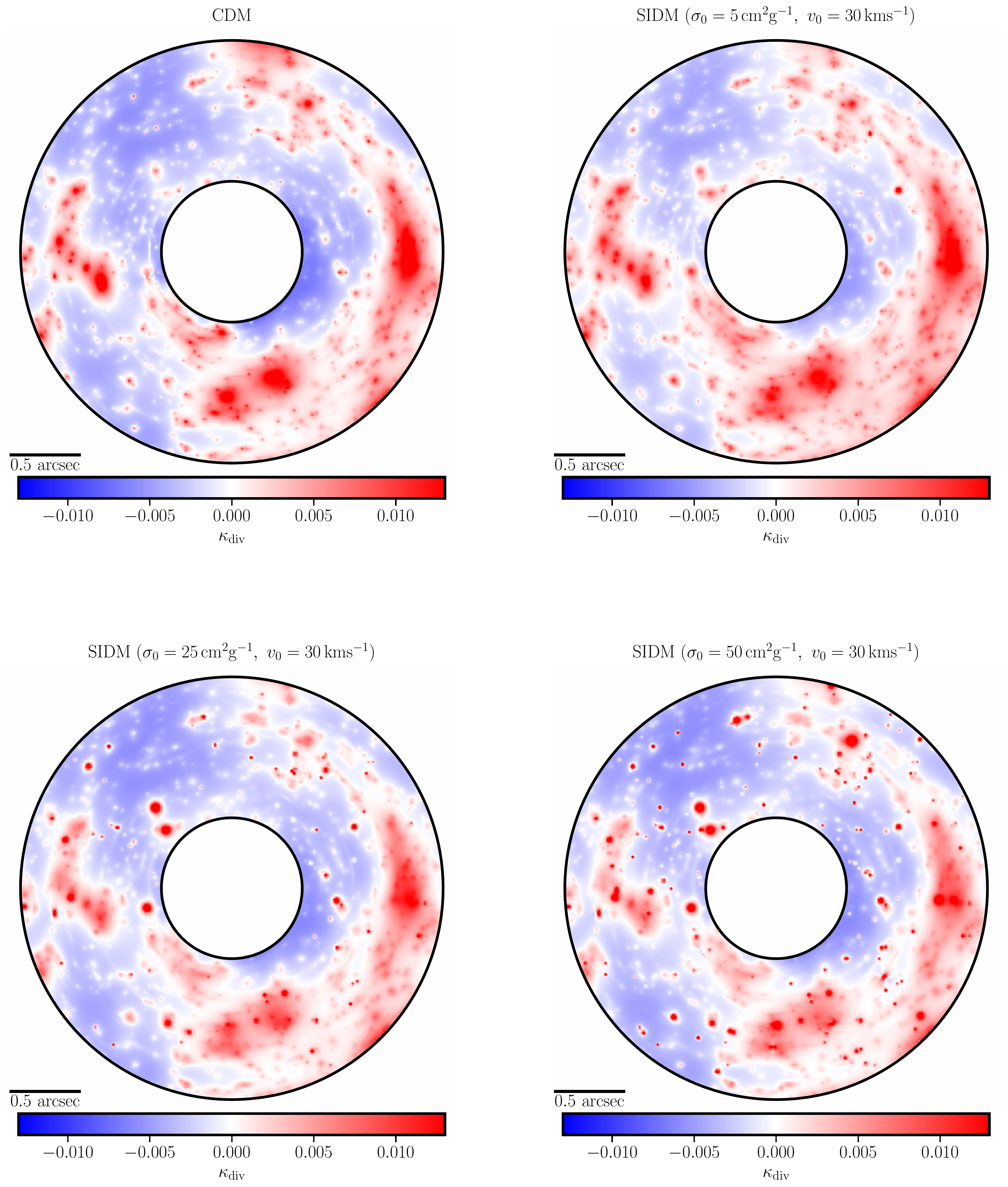}
\end{center}	
\caption{\label{fig:SIDM_CDM_kappa_div} The $\kappa_{\rm div}$ convergence maps for three SIDM models with different interaction cross-sections and a CDM realization take into account both the contribution from the main lens substructure and line-of-sight haloes. As the cross-section increases, the number of core-collapsed subhaloes and line-of-sight haloes increases, resulting in high-density peaks on the convergence maps.}
\end{figure*}

\begin{figure*}
\begin{center}
	\includegraphics[clip, trim=0cm 0cm 0cm 0cm, width=0.8\textwidth]{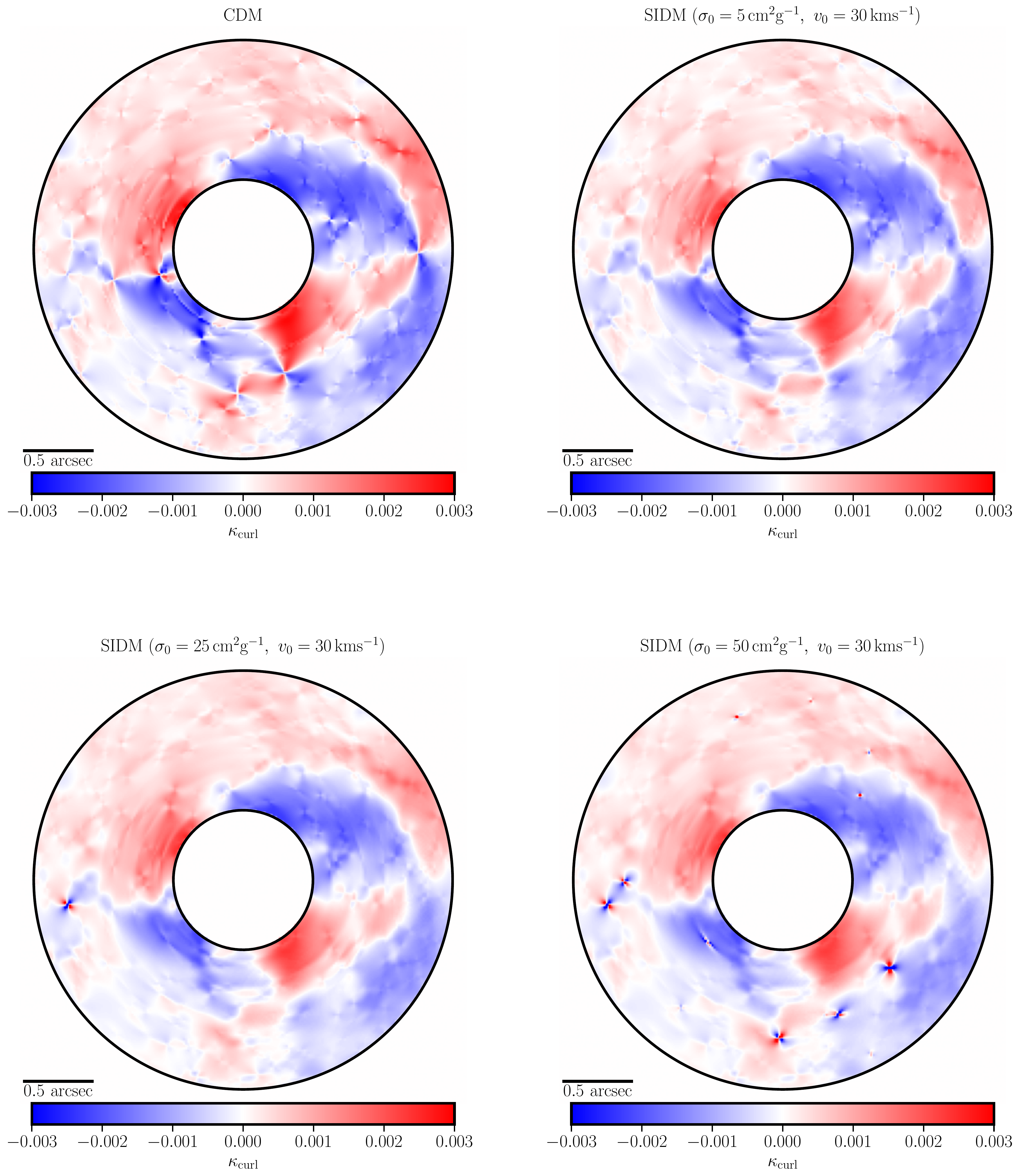}
\end{center}	
\caption{\label{fig:SIDM_CDM_kappa_curl} The $\kappa_{\rm curl}$ convergence maps for SIDM and CDM models shown in Fig.~\ref{fig:SIDM_CDM_kappa_div}. Only line-of-sight haloes contribute to these maps, and steep density cusps of core-collapsed such haloes imprint conspicuous quadrupole patterns in the convergence fields of SIDM realizations.}
\end{figure*}


\subsubsection{Decomposing the two-point function of $\kappa_{\rm curl}$ field}

As mentioned in Section~\ref{Parity_E_and_B}, the quadrupole structures in the $\kappa_{\rm curl}$ field have a unique symmetry structure with odd parity. Therefore, it is useful to introduce a set of orthonormal basis functions, $\{A_{\theta} \sin (\ell \theta)\}$, to decompose the two-point correlation function, $\xi_{\rm curl}(\rr)$, by taking into account the odd-parity of these quadrupole structures in order to extract useful information from the convergence field $\kappa_{\rm curl}$ about the small-scale dark matter structure along the line-of-sight of the observer and the source. Then we can write down the two-point function as:
\be \label{Eq.corr_dec-curl}
\xi_{\rm curl}(\rr)=\xi_{\rm curl}(r, \theta)=\sum_{\ell=1}^\infty \xi_{\rm curl, \ell}(r)A_{\theta} \sin (\ell \theta),
\ee where the parameter

\be 
A_{\theta} =
  \begin{cases}
    +1  & \quad 0 \le \theta < \frac{\pi}{2},\\
   -1  & \quad \frac{\pi}{2} \le \theta < {\pi},
  \end{cases}
\ee is used to extract the useful information hidden in the previously mentioned odd parity structure. The basis functions we introduced here satisfy the following orthogonality condition:

\begin{align}
    \int_0^\pi {\rm d}\theta \,\, A_{\theta} \sin(\ell \theta) \,\, A_{\theta} \sin(m \theta) & = \int_0^\pi {\rm d}\theta \,\,  \sin(\ell \theta) \,\, \sin(m \theta)\en
     & = \frac{\pi}{2} \de_{\ell m},
\end{align} where $\de_{\ell m}$ is the Kronecker delta. Using these relations, the corresponding correlation multipoles of order $\ell$ can be written as
\be 
\xi_{\rm curl, \ell}(r) = \frac{2}{\pi}\int_{0}^\pi {\rm d} \theta \,\, \xi_{\rm curl}(r, \theta)A_{\theta} \sin (\ell \theta).
\ee

\section{RESULTS}\label{Results}
In this section, we will look at the anisotropies that emerge in the $\kappa_{\rm div}$ and $\kappa_{\rm curl}$ maps, focusing on SIDM and WDM theories while utilizing the typical CDM model as a reference. We build our dark matter realizations using the mass function and density profile options described in Section~\ref{Sec: Realizations}.

\subsection{SIDM}\label{SIDM_multipole_sec}

Fig.~\ref{fig:SIDM_CDM_kappa_div} depicts the two-dimensional projected mass density field, $\kappa_{\rm div}$, using three SIDM models with $\sigma_0= 5, \,25, \,50 \, {\rm cm^{2}g^{-1}}$ and $v_0=30 \, {\rm km}{\rm s^{-1}}$, with the CDM model serving as the reference. When the velocity-dependent interaction cross-section is increased, more SIDM substructure and line-of-sight haloes begin to form cores and become less dense. As we further increase the cross-section these haloes begin to core-collapse, resulting in high-density peaks on these maps. The top-right panel of Fig.~\ref{fig:SIDM_CDM_kappa_div} contains a small number of core-collapsed subhaloes and zero core-collapsed line-of-sight haloes, whereas the bottom-right panel contains more core-collapsed line-of-sight dark matter haloes represented by dense structure stretched in the tangential direction, in addition to a relatively larger number of core-collapsed subhaloes appearing as dense circular structure. Fig.~\ref{fig:SIDM_CDM_kappa_curl} illustrates the $\kappa_{\rm curl}$ field of the same SIDM and CDM models. The quadrupolar pattern at the places of line-of-sight haloes distinguishes these maps. When compared to the CDM model and SIDM models with small velocity-dependent cross-sections, the SIDM models with large cross-sections imprint more pronounced quadrupole patterns due to the steep density cusps of core-collapsed line-of-sight dark matter haloes. Furthermore, for large cross-sections, the large-scale features are obviously getting more diffuse, even as the small haloes begin core-collapsing. When compared to CDM haloes with cuspy density profiles, the low central density of cored SIDM substructure and line-of-sight haloes contribute to the convergence maps with faint features in both Figs.~\ref{fig:SIDM_CDM_kappa_div} and \ref{fig:SIDM_CDM_kappa_curl}.

\begin{figure}
\begin{center}
\includegraphics[clip, trim=0cm 0cm 0cm 0cm, width=0.45\textwidth]{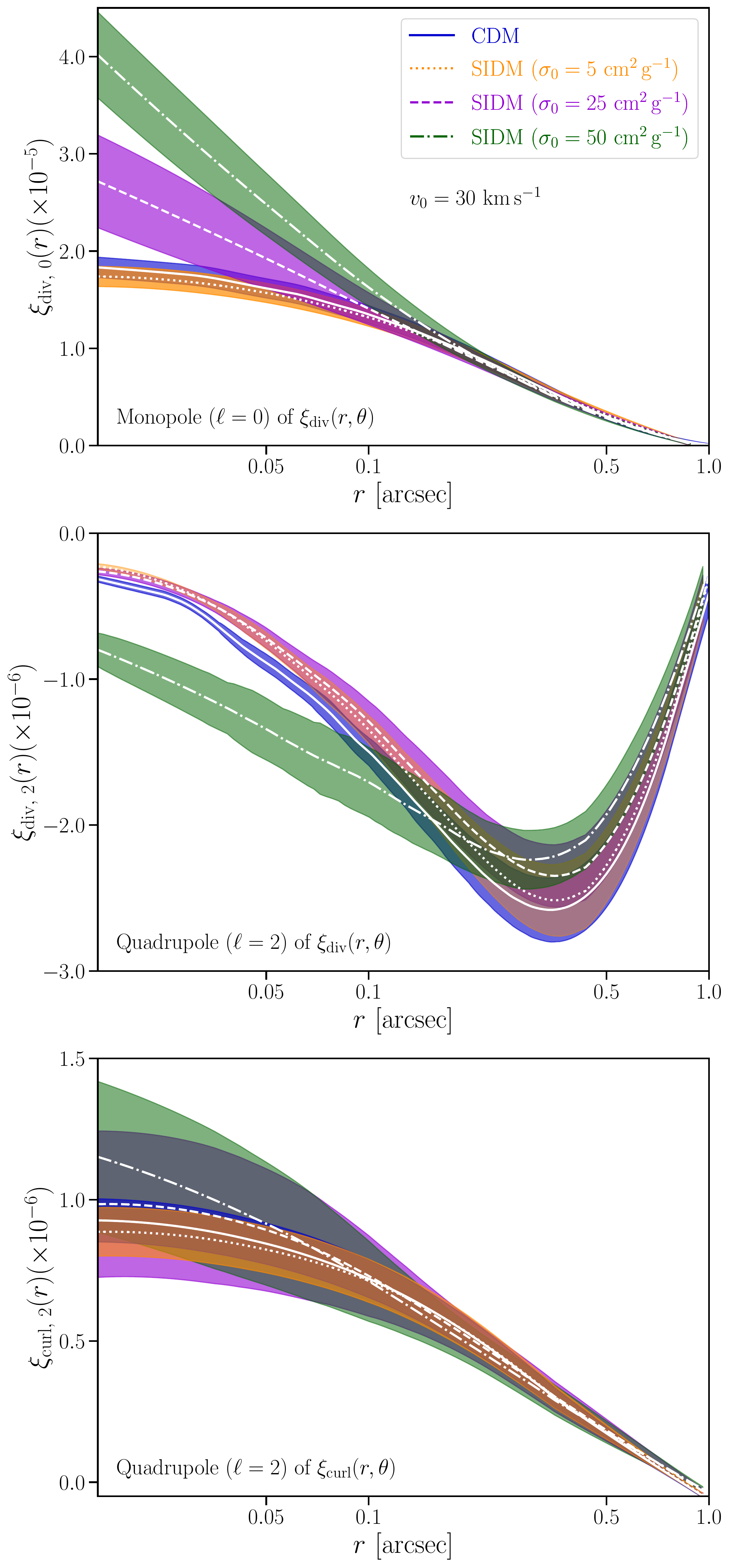}
\end{center}
\caption{\label{SIDM_multipoles_1} The monopole (top panel) and the quadrupole (middle panel) moments of the two-point function $\xi_{\rm div}$, and the quadrupole (bottom panel) moment of $\xi_{\rm curl}$ function  as a function of the velocity-dependent interaction cross-section normalization ($\sigma_0$) for three SIDM models. The CDM model serves as the reference point in this case. All three panels in this narrative have the same legend. The 68\% credible intervals are shown in the shaded areas. The amplitudes of multipoles are primarily determined by the density profile shape and the abundance of cored and core-collapsed haloes.}
\end{figure}

\begin{figure}
\begin{center}
\includegraphics[clip, trim=0cm 0cm 0cm 0cm, width=0.45\textwidth]{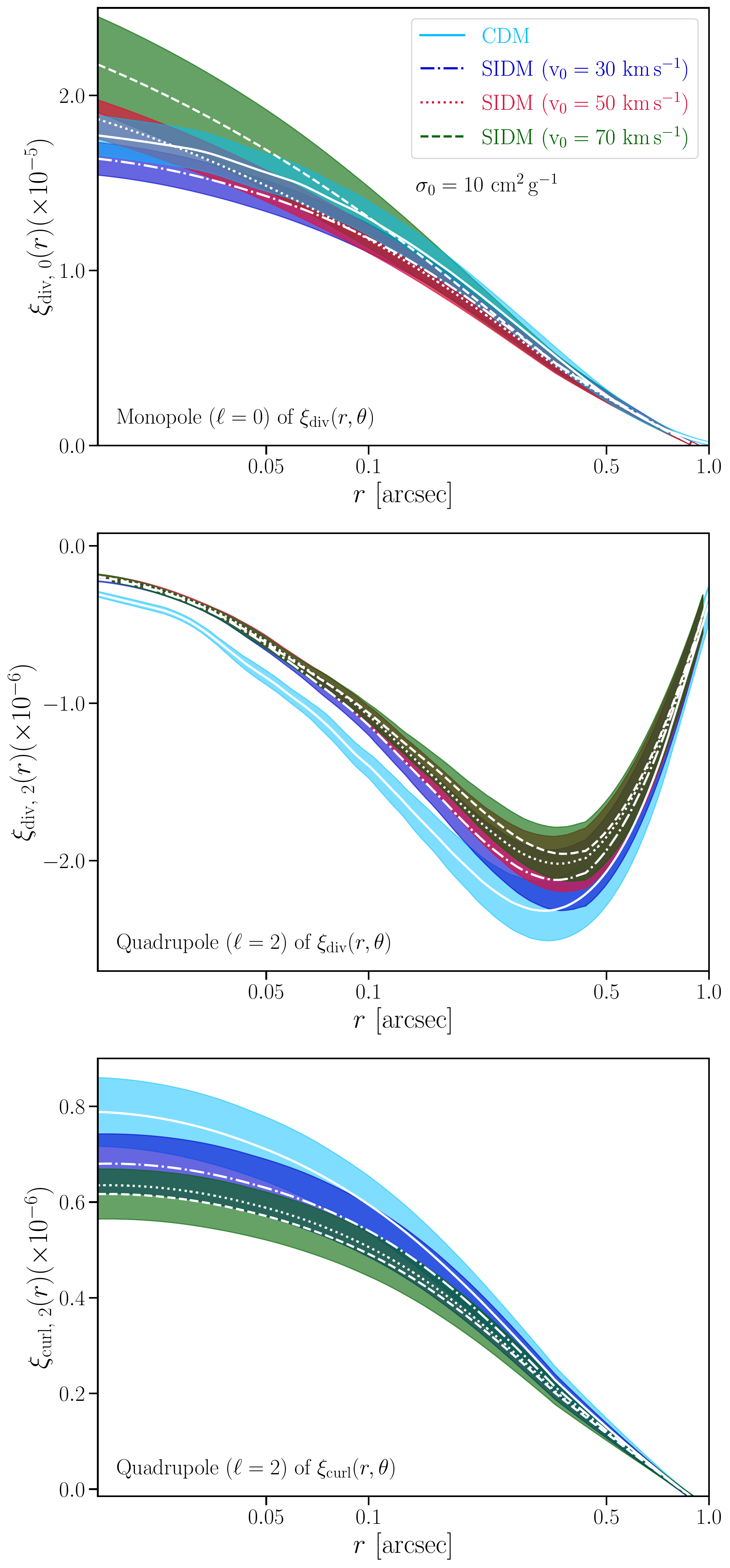}
\end{center}
\caption{\label{SIDM_multipoles_2} The same as in Fig.~\ref{SIDM_multipoles_1}, but this time as a function of the characteristic velocity ($v_0$).}
\end{figure}

The monopole and quadrupole moments of the two-point function, $\xi_{\rm div}$, as well as the quadrupole moments of the two-point correlation function, $\xi_{\rm curl}$, are shown in Fig.~\ref{SIDM_multipoles_1} for three different SIDM models with velocity-dependent self-interaction cross-section normalization ($\sigma_0$) values of 5, 25, 50 $\rm cm^2g^{-1}$ and $v_0 = 30{~\rm kms^{-1}}$, and the CDM model as the benchmark. These plots have been made using 100 different realizations for each dark matter model. The lines reflect the mean, and the dark and light-shaded regions represent 68 and 90 per cent credible intervals, respectively. 
Due to the shallow density cores and the relatively small number of core-collapsed haloes, 
the SIDM model with the smallest cross-section has a lower relative amplitude compared to the CDM model in all three panels of Fig.~\ref{SIDM_multipoles_1}. In comparison, the SIDM model with $\sigma_0=25~{\rm cm^2g^{-1}}$ has a large number of core-collapsed subhaloes and a small number of core-collapsed line-of-sight haloes. 
This core-collapsed substructure significantly contributes to the increase in the amplitude of the monopole of the $\xi_{\rm div}$ function, while the presence of mostly cored line-of-sight haloes slightly decreases its quadrupole moment compared to the CDM case. Meanwhile, the very small number of core-collapsed line-of-sight haloes in this case slightly increases the quadrupole moment of the $\xi_{\rm curl}$ function on small scales. 

The SIDM model with the largest cross-section produces a significant number of core-collapsed line-of-sight haloes in addition to the core-collapsed substructure, which contributes significantly to the large amplitude of the monopole of the $\xi_{\rm div}$ function and the quadrupole of the $\xi_{\rm curl}$ function, which purely represent the line-of-sight haloes. More interestingly, we can see a large relative amplitude of the quadrupole of the $\xi_{\rm div}$ function at small radial distances compared to the SIDM models with small cross-sections due to the significant number of core-collapsed line-of-sight haloes. At small radial distances, this increase in amplitude is accompanied by a steepening of the slope of $\xi_{\rm div,2}(r)$, reflecting the steeper density profile of collapsed line-of-sight haloes. However, at large radial distances, the cored SIDM line-of-sight haloes dominate the amplitude of this quadrupole moment, resulting in a modest relative amplitude compared to the other three models. One of the most apparent features of all three panels is that when moving toward large angular separations, all shown correlation functions for the different models considered asymptote to the same behaviour, echoing the fact that SIDM does not significantly alter the outskirt of dark matter haloes.

The influence of the characteristic velocity, $v_0$, contained in the velocity-dependent cross-section on the multipole moments of the two-point functions $\xi_{\rm div}$ and $\xi_{\rm curl}$ is seen in Fig.~\ref{SIDM_multipoles_2}. As expected, the relative amplitudes of the quadrupole moments of these two-point functions decrease as the interaction cross-section grows. This reflects the presence of a larger core in more massive line-of-sight haloes and subhaloes as $v_0$ is increased. Furthermore, when $v_0$ grows the amplitude of the $\xi_{\rm div}$ monopole initially reduces (reflecting core formation) and subsequently increases in comparison to the benchmark CDM model (reflecting the presence of core-collapsed subhaloes for the $v_0=70 \, {\rm kms^{-1}}$ case).


The relative differences between the two-point correlation function multipoles of the SIDM models we reviewed here and the CDM model are shown in Figs.~\ref{SIDM_multipoles_relative_1} and~\ref{SIDM_multipoles_relative_2} in Appendix~\ref{App. B}. In all, we see that the \emph{combination} of the monopole and quadrupole moments of the $\xi_{\rm div}$ correlation function contains important information about both the amplitude of the self-interaction cross-section at low velocities and about its characteristic velocity. For instance, a steepening of the $\xi_{\rm div,0}$ moment without an accompanied increase in the amplitude of the $\xi_{\rm div,2}$ moment indicates a self-interaction cross-section large enough to core-collapse subhaloes but not line-of-sight haloes, while an increase in both indicates a cross-section large enough to collapse all haloes, whether or not they are in the tidal field of a host. Furthermore, the exact shape of the steepened $\xi_{\rm div,0}$ moment can tell us about the velocity dependence of the cross-section since it reflects which subhalo masses have large cores, which are core-collapsed, and which ones are largely unaffected by self-interaction. 

\subsection{WDM}

\begin{figure}
\begin{center}
\includegraphics[clip, trim=0cm 0cm 0cm 0cm, width=0.45\textwidth]{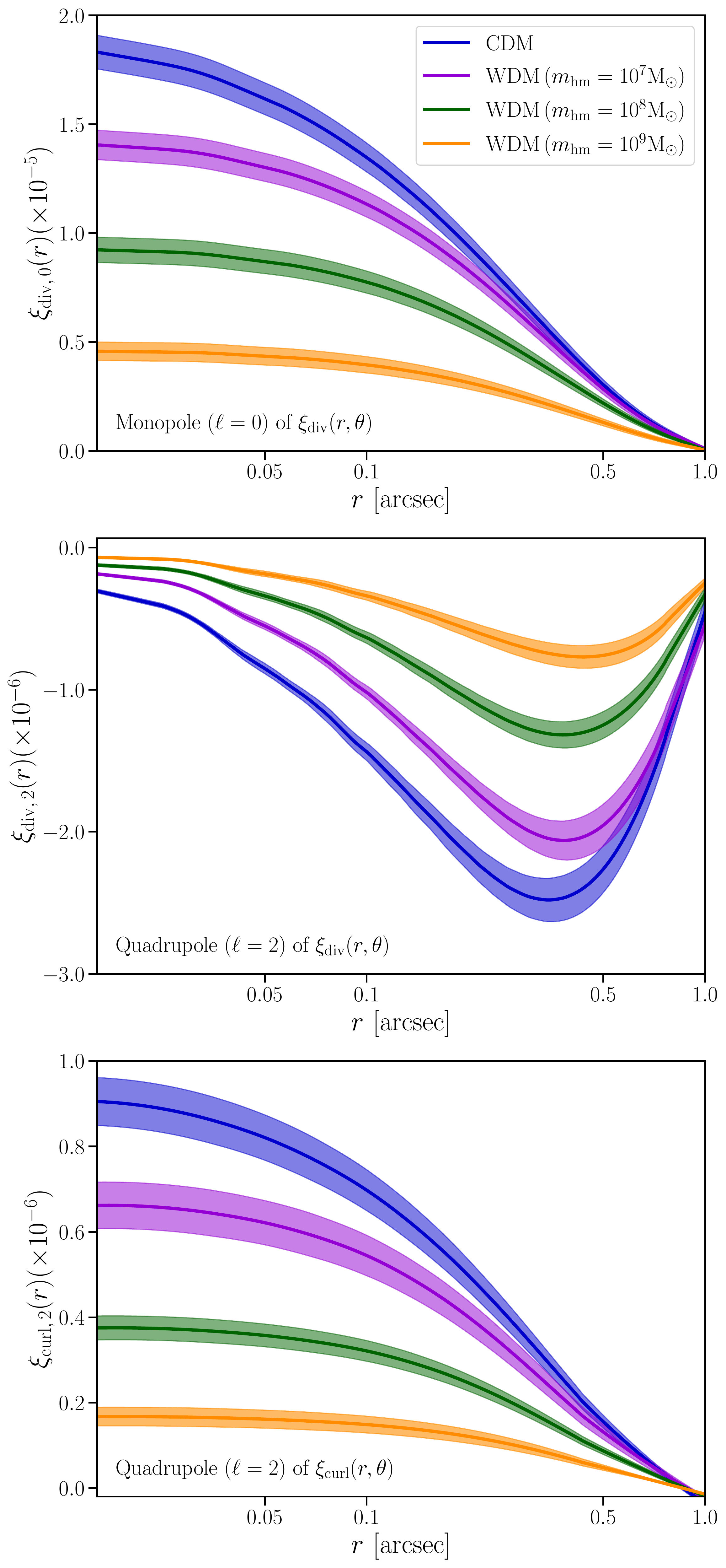}
\end{center}
\caption{\label{WDM_multipoles} The monopole (top panel) and the  quadrupole (middle panel) moments of the two-point function $\xi_{\rm div}$, and the quadrupole (bottom panel) of the function $\xi_{\rm curl}$ as a function of half-mode mass ($m_{\rm hm}$) for three WDM models. The CDM model serves as the benchmark in this case. All three panels in this plot have the same legend. The 68\% credible intervals are shown in the shaded areas. The abundance and concentration of subhaloes and line-of-sight dark matter haloes have the greatest influence on multipole amplitudes.}
\end{figure}

The monopole and quadrupole moments of the two-point function $\xi_{\rm div}$, as well as the quadrupole moment of the $\xi_{\rm curl}$ function, are shown in Fig.~\ref{WDM_multipoles} for three different WDM models with various half-mode masses ($m_{\rm hm}$). The WDM particle mass ($m_{\rm WDM}\sim4.5$ keV) obtained by \cite{Villasenor_2023} using Lyman-$\alpha$ flux power spectrum data corresponds to $m_{\rm hm}\simeq 10^8 {\rm M}_\odot$ line in Fig.~\ref{WDM_multipoles}. As the half-mode mass increases, the number of substructure and line-of-sight haloes decreases, and the central densities of such WDM haloes are relatively suppressed compared to CDM haloes of the same mass due to their lower concentrations. Because the amplitudes of two-point function multipoles are sensitive to the central density of dark matter haloes, particularly at small radial distances, the suppression of halo concentrations for masses below $m_{\rm hm}$ produces smaller perturbations to the lensed images than CDM haloes, and thus the anisotropic signal imprinted by line-of-sight dark matter haloes is low. Therefore these effects can be seen as a progressive decrease in the relative amplitude of the multipoles with respect to the CDM model as the half-mode mass increases. 

For small radial distances, the amplitude of the different multipole moments diminishes as the half-mode mass grows. This reflects the suppressed abundance of small haloes, which are the primary contributors to the correlation function at these scales. At large radial distances, the correlation functions for the different models shown converge to the same behaviour because the abundance of large haloes (which dominate at these scales) is essentially unaffected by these free-streaming effects.

\section{DISCUSSION AND Conclusions}\label{Conclusions}
In this work, we have investigated how the two-point correlation function of the $\kappa_{\rm div}$ field of dark matter haloes in a strong lens system evolves when both self-interacting and warm dark matter scenarios are considered while maintaining the cold dark matter framework as the benchmark. We were particularly interested in the parity-even quadrupole moment of this two-point function, which is produced by the anisotropic fingerprints generated by line-of-sight dark matter haloes due to non-linear coupling between various lens planes. We also developed a framework for computing the parity-odd quadrupole moment of the $\kappa_{\rm curl}$ field's two-point function, which only arises in the presence of line-of-sight dark matter haloes. We investigated how the correlation function multipoles behave as a function of the self-interaction cross-section in the SIDM scenario, which essentially dictates the abundance and density profiles of cored or core-collapsed haloes in a strong lens system. Our results can be summarized as follows:
 
\begin{itemize}
  \item For small values of the cross-section, cored haloes dominate the amplitudes of the monopole and quadrupole of the two-point function of the $\kappa_{\rm div}$ field, resulting in a reduced relative amplitude compared to the CDM scenario.

  \item As the self-interaction cross-section grows, subhaloes can begin to core-collapse, resulting in a steeper and larger monopole at small radial distances. Meanwhile, line-of-sight haloes, which are not subjected to tidal truncation, are still in the cored density profile regime, resulting in a relatively suppressed quadrupole moment of the $\kappa_{\rm div}$ field as compared to the CDM case. 

  \item As the self-interaction cross-section grows further, both subhaloes and line-of-sight haloes undergo core collapse, resulting in both the monopole and quadrupole of the $\xi_{\rm div}$ correlation function becoming large and steep at small radial distances. In parallel, the parity-odd (curl) quadrupole correlation function displays a slight increase in amplitude at small radial distances due to the core-collapsed line-of-sight haloes. 

  \item For large enough self-interaction cross-section, the exact slope of the monopole and quadrupole of the $\xi_{\rm div}$ correlation function at small radial distances depends on the characteristic velocity $v_0$.

\end{itemize}

\emph{It is thus clear that using detailed observations with high signal-to-noise ratio and resolution, both the monopole and quadrupole moments of the $\xi_{\rm div}$ correlation function, especially at small radial distances, could constrain both the amplitude and velocity dependence of the SIDM cross-section.}







 In the WDM scenario, we investigated the change in amplitudes of the two-point function multipoles as a function of the WDM's characteristic half-mode mass, which controls the abundance of small dark matter haloes. In this case, the relative amplitudes of the different multipoles are mostly determined by the abundance of low-mass haloes. As a result, when the half-mode mass increases, the number of small haloes reduces, as do the amplitudes of the multipole moments at small radial separation.
 
 In conclusion, based on the shapes and amplitudes of the two-point function multipoles in the two different dark matter scenarios we considered, our work shows that extracting these multipole moment signals from galaxy--galaxy strong lens observations is a promising technique for studying dark matter physics. However, a detailed study is needed to determine how well this anisotropic signal could be extracted from realistic strong lenses. Encouragingly, \cite{Dhanasingham_2022} showed that both the monopole and quadrupole of  the $\xi_{\rm div}$ correlation function could be jointly extracted from high-resolution mock observations of a galaxy-scale Einstein ring. Future work is needed to assess the power of a large sample of strong lenses to constrain the SIDM cross-section and the abundance of small-mass haloes via the key anisotropic signatures we have pointed out in this work.  With upcoming space- and ground-based surveys in this decade to observe and detect strong lens systems \citep{Serjeant_2014, Collett_2015, Serjeant_2017, Metcalf_2019, Weiner_2020, Mao_2022}, such a study will help us better grasp the dark Universe.
 
 
\section*{Acknowledgements}

We would like to express our gratitude to the anonymous referee for providing insightful comments and suggestions on this manuscript for improvements. We would also like to thank Tansu Daylan, Daniel Gilman, Xiaolong Du, Carton Zeng, Ekapob Kulchoakrungsun, Kylar L. Greene, David Camarena, and Soumyodipta Karmakar for useful discussions. B.D. and F.-Y.C.-R. acknowledge the support of program HST-AR-17061.001-A whose support was provided by the National Aeronautical and Space Administration (NASA) through a grant from the Space Telescope Science Institute, which is operated by the Association of Universities for Research in Astronomy, Incorporated, under NASA contract NAS5-26555. This work was supported in part by the NASA Astrophysics Theory Program under grant 80NSSC18K1014. F.-Y. C.-R. would like to thank the Robert E. Young Origins of the Universe Chair fund for its generous support. We also would like to thank the UNM Center for Advanced Research Computing, supported in part by the National Science Foundation, for providing the high-performance computing resources used in this work. B. D. would like to thank Kevin Fotso Tagne and Bhashithe Abeysinghe for their assistance with high-performance computing. The work in this manuscript made partial use of the \textsc {python} packages \textsc {numpy} \citep{Numpy}, \textsc {numba} \citep{numba}, \textsc {matplotlib} \citep{matplotlib}, \textsc {scipy} \citep{SciPy}, \textsc {astropy} \citep{astropy}, and \textsc {colossus} \citep{COLOSSUS}. We are appreciative of the great work done by the developers of these software.

\section*{Data Availability}

All the data generated in this research, as well as the codes used, will be available upon reasonable request to the corresponding author.



\bibliographystyle{mnras}
\bibliography{references} 




\appendix

\section{Determining the Core density}\label{App. A}

In this appendix, we present briefly the process for calculating the value of the core radius to scale radius ratio, as stated in \cite{gilman2021strong}.

Rewrite the equation~\eqref{ctnfw}.

\begin{equation}
    \rho(x, \beta, \tau)=\frac{\rho_{\rm s}}{(x^a+\beta^a)^\frac{1}{a}(1+x)^2}\frac{\tau^2}{\tau^2+x^2}
\end{equation} If we set the core density $\rho_0$ to $\rho(0, \beta, \tau)=\rho_0$, we get

\be
\beta = \frac{r_{\rm c}}{r_{\rm s}} = \frac{\rho_{\rm s}}{\rho_{\rm 0}},
\ee which determines the core radius $r_{\rm c}$. \cite{gilman2021strong} used a simple Jeans modelling approach previously proposed by \cite{Kaplinghat:2015aga} to compute the core density $\rho_0$ and velocity dispersion $v_{\rm rms}$. Dark matter particles scatter more efficiently near the centre of the halo, where the density of dark matter particles is greatest. As a result, \cite{Kaplinghat:2015aga} divide the halo into two regions delineated by a distinctive radius $r_1$, where this radius $r_1$ meets the condition:
\be 
\rho_{\rm NFW}(r_1)\left<\sigma v \right>t_{\rm halo}=1,
\ee where $\left<\sigma v \right>$ describes the scattering rate given the cross-section $\sigma$ and the relative velocity $v$, and $\rho_{\rm NFW}$ is the density of the NFW \citep{NFW_1996} profile. At radius, $r>r_1$, \cite{Kaplinghat:2015aga} assume that the density profile acts like an NFW profile, which is similar to collisionless CDM particles due to low dark matter self-interactions. By assuming that haloes collapse at redshift $z = 10$, we calculate the halo age $t_{\rm halo}$ as a function of $z$.
The density profile of the isothermal solution $\rho_{\rm iso}$ is determined using the radial Jeans equation, 
\be
v_{\rm rms}^2 \bna \rho_{\rm iso}=-\rho_{\rm iso} \bna \Phi
\ee assuming hydrostatic equilibrium due to dark matter self-interactions. The differential equation for $\rho_{\rm iso}$ can be expressed using Poisson's equation $\nabla^2 \Phi=4\pi G \rho_{\rm iso}$ followed by the total gravitational potential $\Phi$ from dark matter as
\be 
v_{\rm rms}^2 \nabla ^2 \ln \rho_{\rm iso}=-4\pi G\rho_{\rm iso}.
\ee This differential equation of the model with two unknowns, $v_{\rm rms}$ and $\rho_{0}$, is solved using the boundary conditions $\rho_{\rm iso}(0)=\rho_0$ and $\rho ' _{\rm iso}(0)=0$. Since baryons are subdominant inside dark matter haloes in the mass range we consider in our simulations, we ignore their contribution to the mass of the dark matter haloes while solving these differential equations. According to \cite{gilman2021strong}, \textsc {pyhalo} employs a grid-based search approach to identify the best-fitting values for $\rho_{0}$ and $v_{\rm rms}$ until it meets the following two conditions at $r_1$ to 1\% precision. The first condition states that the amplitude of the density of the isothermal solution $\rho_{\rm iso}$ at $r_1$ matches that of the NFW profile $\rho_{\rm NFW}$ at $r_1$, and the second condition states that the enclosed mass of the isothermal solution within $r_1$ must equal that of the NFW profile with no interactions.

\section{Relative differences between SIDM and CDM multipoles}\label{App. B}

\begin{figure}
\begin{center}
\includegraphics[clip, trim=0cm 0cm 0cm 0cm, width=0.456\textwidth]{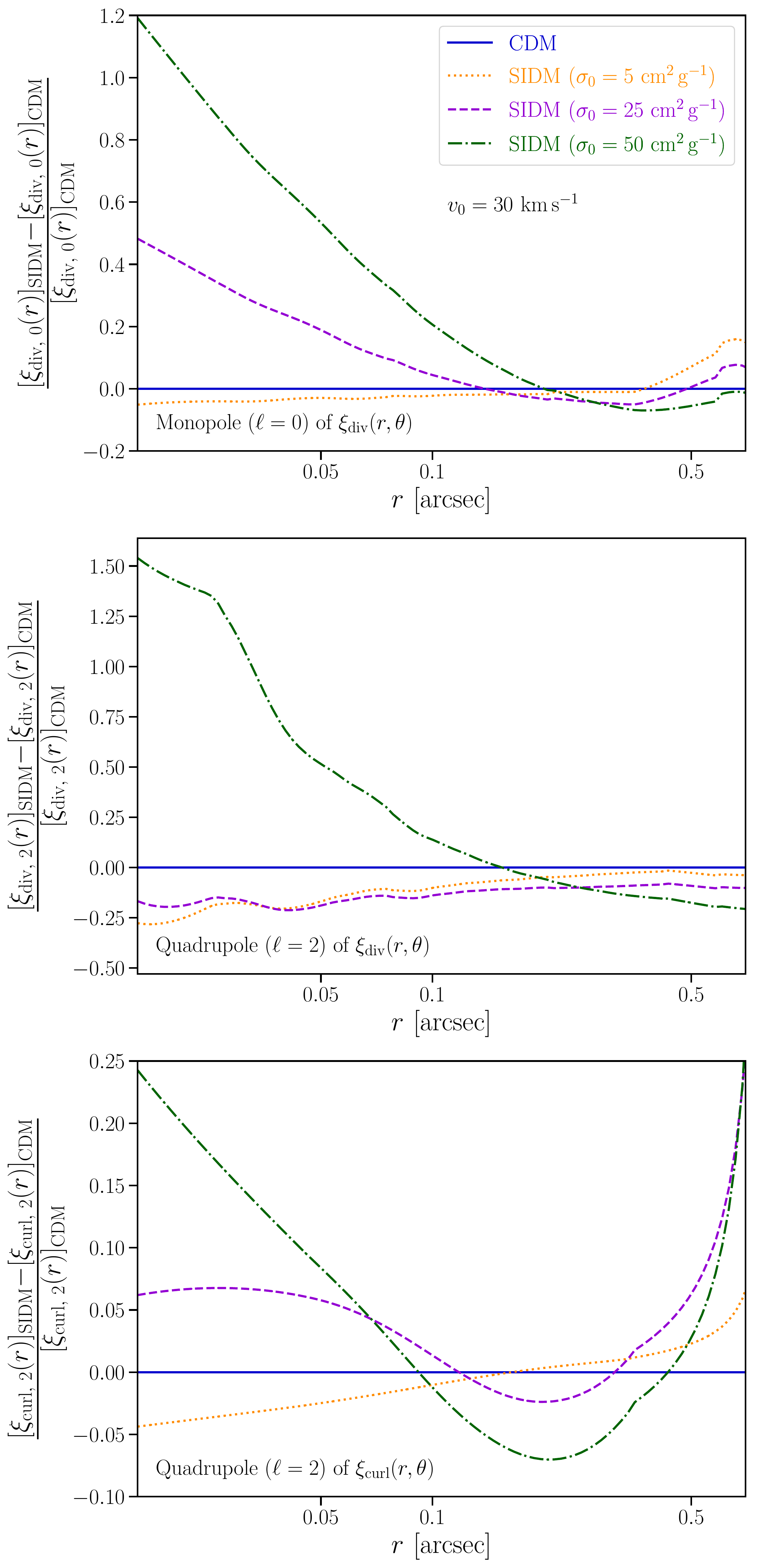}
\end{center}
\caption{\label{SIDM_multipoles_relative_1} The relative difference between the two-point function multipoles (monopole (top panel) and the quadrupole (middle panel) of $\xi_{\rm div}$, and the quadrupole moment (bottom panel) of $\xi_{\rm curl}$) of the three SIDM models shown in Fig.~\ref{SIDM_multipoles_1} and the CDM model as a function of the velocity-dependent interaction cross-section normalization ($\sigma_0$). All three panels in this narrative have the same legend.}
\end{figure}

\begin{figure}
\begin{center}
\includegraphics[clip, trim=0cm 0cm 0cm 0cm, width=0.456\textwidth]{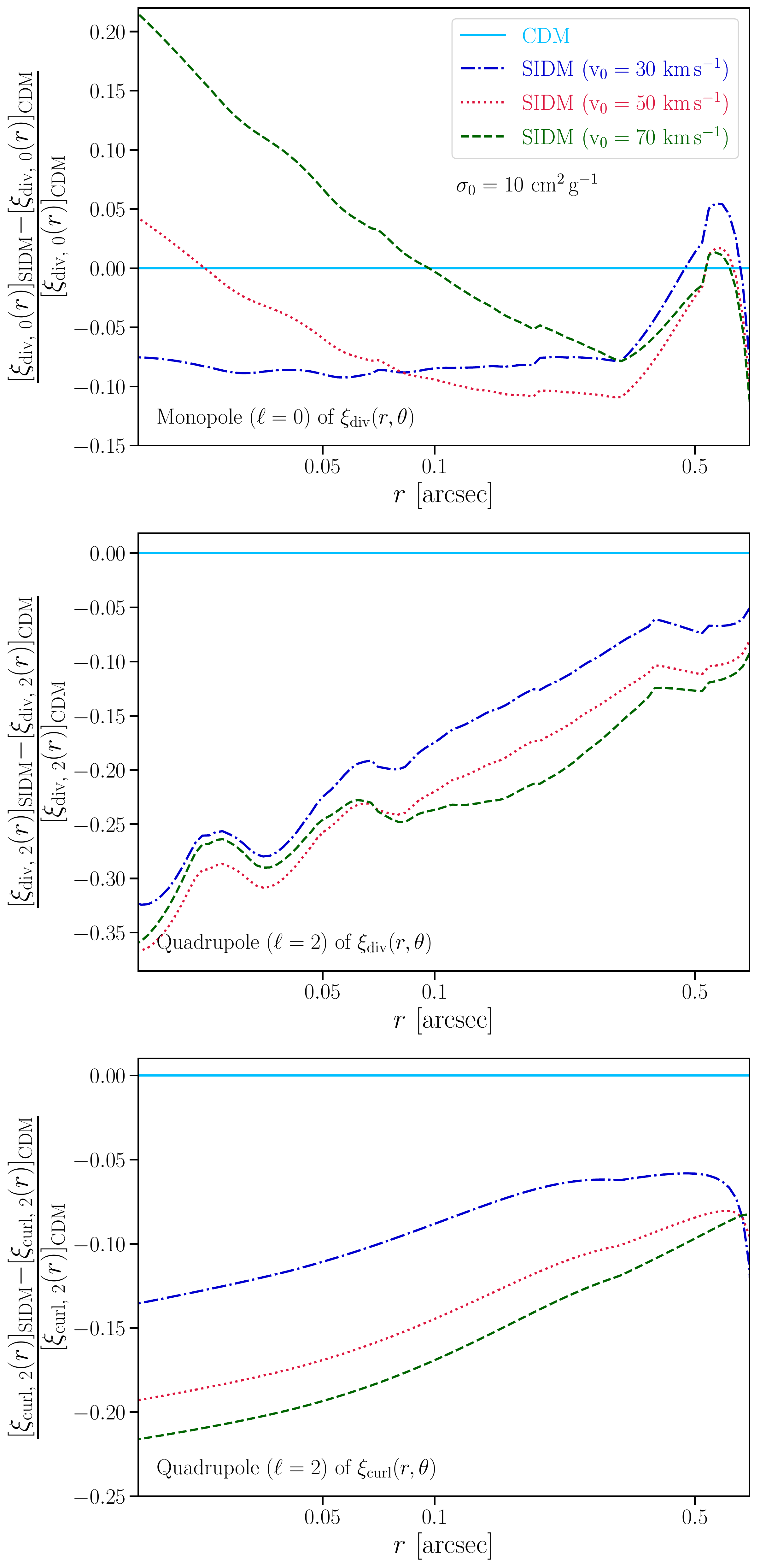}
\end{center}
\caption{\label{SIDM_multipoles_relative_2} The same as in Fig.~\ref{SIDM_multipoles_relative_1}, but now as a function of the characteristic velocity ($v_0$) for the SIDM models depicted in Fig.~\ref{SIDM_multipoles_2}.}
\end{figure}

In this appendix, the relative differences between the CDM and the SIDM two-point correlation multipoles discussed in Section~\ref{SIDM_multipole_sec} are shown in Figs.~\ref{SIDM_multipoles_relative_1} and \ref{SIDM_multipoles_relative_2}.
Also, Table~\ref{Tab. B1} displays the mean number of subhaloes, line-of-sight haloes, and core-collapsed haloes, as well as their fractions, as a function of the SIDM cross-section, to offer quantitative information regarding core collapse.

\begin{table} \label{Tab. B1}
	\centering
	\caption{The mean number of haloes and core-collapsed haloes in our SIDM realizations used to obtain the results shown in Figs.~\ref{SIDM_multipoles_1} and \ref{SIDM_multipoles_relative_1}. Notations $\left< N_{\rm SH}\right>$, $\left< N_{\rm LOSH}\right>$, $\left< N_{\rm cc, SH}\right>$, and $\left< N_{\rm cc, LOSH}\right>$ represent the mean numbers of subhaloes, line-of-sight haloes, core-collapsed subhaloes, and core-collapsed line-of-sight haloes, respectively.}
	\label{tab:example_table}
	\begin{tabular}{lccr} 
		\hline
		$v_0$ [$\rm {km s^{-1}}$] & \multicolumn{3}{c}{\centering 30}\\
		\hline
        $\sigma_0$ [$\rm {cm^2 g^{-1}}$] & 5 & 25 & 50\\
        \hline
        $\left< N_{\rm SH}\right>$ & \multicolumn{3}{c}{\centering 2032} \\
        $\left< N_{\rm LOSH}\right>$ & \multicolumn{3}{c}{\centering 1575}\\  
		$\left< N_{\rm cc, SH}\right>$  &  4 & 408 &  833\\ 
        $\frac{\left< N_{\rm cc, SH}\right>}{\left< N_{\rm SH}\right>}$ & 0.002 & 0.201 & 0.410\\
		$\left< N_{\rm cc, LOSH}\right>$  & 0 & 2 & 62\\
        $\frac{\left< N_{\rm cc, LOSH}\right>}{\left< N_{\rm LOSH}\right>}$ & 0.0 & 0.001 & 0.039\\
		\hline
	\end{tabular}
\end{table}


\bsp	
\label{lastpage}
\end{document}